\documentclass{aa}  

\usepackage{graphicx}
\usepackage{amsbsy}
\usepackage{txfonts}
\usepackage{hyperref}
\hypersetup{
       hyperfootnotes=true,                     
           bookmarks=true,                      
           colorlinks=true,
           linkcolor=black,
            linktoc=page,
           anchorcolor=black,
           citecolor=blue,
           urlcolor=black,
}
\DeclareUnicodeCharacter{2212}{-}
\begin{document}

   \title{Feeding and feedback processes in the Spiderweb proto-intracluster medium}


   \author{M. Lepore
          \inst{1}
           \and
          L. Di Mascolo
          \inst{2,3,4}
          \and
          P. Tozzi
          \inst{1}
          \and
          E. Churazov
          \inst{5}
          \and
          T. Mroczkowski
          \inst{6}
          \and
          S. Borgani
          \inst{2,3,4,7}
          \and
          C. Carilli
          \inst{8}
          \and
          M.Gaspari
          \inst{9}
          \and
           M. Ginolfi
          \inst{10}
          \and
          A. Liu
          \inst{11}
          \and
          L. Pentericci
          \inst{12}
          \and
          E. Rasia
          \inst{3,4}
          \and
          P. Rosati
          \inst{13}
          \and
          H.J.A. R\"ottgering
          \inst{8}
          \and
          C. S. Anderson
          \inst{14}
          \and
          H. Dannerbauer
          \inst{15,16}
          \and
          G. Miley
          \inst{17}
          \and
          C. Norman
          \inst{18,19}
          }

\institute{INAF-Osservatorio Astrofisico di Arcetri, Largo Enrico Fermi 5, 50125, Florence, Italy
\and Astronomy Unit, Department of Physics, University of Trieste, via Tiepolo 11, I-34131 Trieste, Italy
\and INAF-Osservatorio Astronomico di Trieste, via G. B. Tiepolo 11, I-34143 Trieste, Italy
\and IFPU - Institute for Fundamental Physics of the Universe, Via Beirut 2, 34014 Trieste, Italy
\and Max Planck Institute for Astrophysics, 
Karl-Schwarzschild-Str. 1, D-85741 Garching, Germany
\and European Southern Observatory (ESO), Karl-Schwarzschild-Str. 2, D-85748 Garching, Germany
\and INFN–Sezione di Trieste, Trieste, Italy
\and National Radio Astronomy Observatory, P. O. Box 0, Socorro, NM 87801, USA
\and Department of Astrophysical Sciences, Princeton University, Princeton, NJ 08544, USA
\and Department of Physics and Astronomy, University of Florence, Via Giovanni Sansone, 1, 50019 Florence, Italy
\and Max Planck Institute for Extraterrestrial Physics, Giessenbachstrasse 1, 85748 Garching, Germany 
\and INAF - Osservatorio Astronomico di Roma, Via Frascati 33, I-00040 Monteporzio (RM), Italy
\and Dipartimento di Fisica e Scienze della Terra, Università degli Studi di Ferrara, via Saragat 1, I-44122 Ferrara, Italy
\and Jansky Fellow of the National Radio Astronomy Observatory, P. O. Box 0, Socorro, NM 87801, USA 
\and Instituto de Astrofísica de Canarias (IAC), 38205 La Laguna, Tenerife, Spain
\and Universidad de La Laguna, Dpto. Astrofísica, 38206 La Laguna, Tenerife, Spain
\and Leiden Observatory, PO Box 9513, 2300 RA Leiden, The Netherlands
\and Space Telescope Science Institute, 3700 San Martin Dr., Baltimore, MD 21210, USA
\and Johns Hopkins University, 3400 N. Charles Street, Baltimore, MD 21218, USA
 }


 
  \abstract
   {We present a detailed analysis of the thermal, diffuse emission of the proto-intracluster medium (proto-ICM) 
   detected in the halo of the Spiderweb Galaxy at $z=2.16,$ within a radius of $\sim$ 150 kpc.}
   {Our main goal is to derive the thermodynamic profiles 
   of the proto-ICM, establish the potential presence
   of a cool core and constrain the classical mass deposition rate (MDR) that may feed the nuclear 
   and the star formation (SF) activity, and estimate the available energy budget 
   of the ongoing feedback process.}
   {We combined deep X-ray data from {\sl Chandra} and 
   millimeter observations of the Sunyaev-Zeldovich (SZ) effect obtained by the Atacama Large Millimeter/submillimeter Array (ALMA)}
   {Thanks to independent measurements of the pressure profile from the ALMA SZ observation and the electron 
   density profile from the available X-ray data, we derived, for the first time, the temperature 
   profile in the ICM of a $z>2$ protocluster. It reveals the presence of a strong 
   cool core (comparable to local ones) that may host a significant mass deposition flow,{ consistent with the measured local SF values}.  
   We also{ find mild evidence of an asymmetry in the X-ray surface brightness distribution, which may be tentatively
   associated with a cavity carved into the proto-ICM by the radio jets. In this case, the estimated 
   average feedback power would be in excess of $\sim 10^{43}$ erg/s. Alternatively, the asymmetry 
   may be due to the young dynamical status of the halo.} }
   {The cooling time of baryons in the core of the Spiderweb Protocluster is estimated to be 
   $\sim 0.1$ Gyr,{ implying that the baryon cycle in the first stages of protocluster formation is characterized by a high-duty cycle and a very active environment}. In the case of the Spiderweb protocluster, 
   we are witnessing the presence of a strongly peaked core that is possibly hosting a cooling
   flow with a MDR up to 250-1000 $\rm M_\odot$/year, 
   responsible for feeding both the central supermassive black hole (SMBH) and the 
   high star formation rate (SFR) observed in the Spiderweb Galaxy. This phase 
   is expected to be rapidly followed by active galactic nucleus (AGN) feedback events, whose onset may 
   have already left an imprint in the radio and X-ray appearance of the Spiderweb protocluster, 
   eventually driving the ICM into a self-regulated, 
   long-term evolution in less than one Gyr. }


   \maketitle
%

\section{Introduction}
\label{sec:Introduction}

The progenitors of local galaxy clusters,  so-called protoclusters, 
are rare overdensities at high-redshift ($z>1.5$) that have not become virialized yet, but are 
expected to reach a  
total virial mass in the range of $10^{14}-10^{15} \, \rm M_\odot$ by $z=0$ \citep{2016Overzier}. 
Observations of these large structures, which span breadths of 10$\arcmin$-30$\arcmin$ 
in the sky \citep{Muldrew2015}, 
are crucial to investigating the formation of large-scale structures in the Universe, particularly in high-density environments.  They will also  improve our understanding of the processes of galaxy formation and evolution, the fueling of star formation (SF) and active galactic nuclei (AGNs), as well as the formation of brightest cluster galaxies (BCGs) and the origin and evolution of 
the ICM, which is the main baryonic component in galaxy clusters. 

\indent 
To date, there has been no consistent, uniform way to search for protoclusters, since they do not reach
a high contrast against the foreground and background extragalactic sky across their entire extent. 
A convenient (albeit potentially biased) technique is to identify high-redshift, massive galaxies 
inhabiting a local overdensity and/or a very active environment, 
as a tracer of the full-scale overdensity.  
In this respect, protocluster search techniques do not differ much from the 
classic methods used in the search for high-redshift massive clusters, such as 
identifying concentrations of massive red sequence galaxies \citep{Bell2004} 
using near-infrared (NIR) and mid-infrared (MIR) filters 
\citep[see][]{Gladders2000,Gladders2005,Wilson2009}; 
searching for extended sources in X-ray surveys \citep{Rosati2002,2022Liu} 
or for the SZ effect associated with the ICM 
\citep{Carlstrom2011,2015Bleem,2018Mantz,2019Gobat};
investigating the environment around powerful high-redshift radio galaxies \citep{Miley2008}. 
In general, for a given protocluster, the average galaxy overdensity 
on scales of several physical Mpc (possibly comprising the whole region that 
eventually will collapse and virialize by $z=0$) may be orders 
of magnitude lower than in virialized clusters.
As a consequence, the majority of the diffuse baryons responsible for the 
X-ray and SZ signal may have not reached a temperature and density sufficient for 
such systems to be detected with current facilities. 
Therefore, while in the case of virialized clusters, a single-band observable 
may be sufficient to efficiently select them and provide a mass scale at the same time, 
the robust identification and characterization of protoclusters necessarily require 
a multiwavelength approach.

\indent 
An unbiased method to identify protoclusters would be the full mapping of the 
member galaxies across several physical Mpc, but this is clearly 
extremely time-consuming and not feasible with current observing methods. While the
situation is likely to improve with the advent of future large-area surveys in the 
optical and NIR, such as the Euclid mission \citep{2012Laureijs} or Rubin Observatory \citep{2019Ivezic}, we have
to rely on biased tracers, as previously mentioned. In this respect, 
enhanced SF and nuclear activity spread across the member galaxies
is expected to be a distinctive feature of high-density environments at high redshift.
In this scenario, X-ray observations play an important role in studying 
the unresolved emission of AGNs and, hence, the accretion phenomena onto SMBHs, emission of star-forming galaxies (SFGs), and inverse Compton (IC) emission from the
relativistic plasma of the radio jets when present.
Furthermore, the X-ray band is key to trace and characterize the proto-ICM 
that is expected to form and appear during the most intense evolutionary 
phase of protoclusters in the range $2<z<3$.
In particular, it is key to understanding whether the forming proto-ICM reaches densities large enough to 
give rise to a cooling flow that, in turn, can feed the process of SF 
in the central halo, usually identified as the BCG undergoing formation.
This occurrence seems to be suggested by the high star formation rates (SFRs) measured in proto-BCGs, 
but the frequency of the cooling flow regime, as well as precipitation and chaotic cold accretion processes (CCA; see \citealp{2020Gaspari} for a review), at $z\geq 2,$ have not been measured yet. 
Another assumption required to strengthen the cooling flow hypothesis at high redshift is that 
in the very first stages of formation of the main halo, 
the feedback processes may not have been triggered yet. 
This would leave room for a significant MDR,
at variance with what is commonly observed during the secular evolution of
massive clusters at $z<1$, where a fully developed and massive cooling flow with a high and 
persistent mass-deposition rate has not been observed \citep{2006Peterson}.
On the other hand, several processes that may potentially heat the proto-ICM, 
thereby hampering its cooling, are also 
occurring on a very short timescales, such as 
intense SF, merging of 
comparable-mass subclusters and accretion shocks,
as well as AGN jets and associated feedback processes, such as a cascading turbulence (e.g., \citealp{2018Gaspari,2021Wittor}). In this framework, 
tracing the proto-ICM in the range of $2<z<3$ would provide a wealth of information on 
a unique and crucial phase of cluster evolution, when  
many independent physical processes that counterbalance each other 
occur at very high intensity and on a short timescale. 

\indent Needless to say, the X-ray emission from the proto-ICM is expected to be very faint.
Models and simulations tell us that at $z>2$ the emission of the proto-ICM is 
characterized by a very low surface brightness
with a small ($\sim 10$ arcsec) extension \citep{2009Saro}. This not only requires
high sensitivity, but also high angular resolution ($\leq 1$ arcsec), 
a feature that is currently provided only by the {\sl Chandra} satellite.
In addition, in the likely case of a protocluster beaconed by a radio galaxy, 
the thermal emission can be overcome by the 
non-thermal emission associated with IC scattering of 
cosmic microwave background (CMB) photons by the 
relativistic electrons of radio jets. \\
\indent The only alternative window to observe the proto-ICM 
with high angular resolution is provided by the SZ effect 
\citep{1972Sunyaev,1980Sunyaev}. 
In particular, the thermal SZ effect directly traces the line-of-sight 
integrated pressure of the free electrons in the diffuse hot baryons. 
As such, it provides a complementary, yet independent view of the same ICM that can be 
imaged in emission 
in the X-ray band \citep[for a review see][]{2019Mroczkowski}. 
In particular, we note that X-ray data are sensitive mostly to the 
square of the electron density, while the SZ signal is directly proportional 
to the pressure; thus, it depends linearly on the electron density.
A relevant point is that both the X-ray flux and the SZ signal from the ICM at a given virial mass at high-z do not depend strongly on redshift \citep{2015Churazov} and, therefore,
both windows are well suited to detect the presence of a pressurized plasma 
out to $z\gtrsim 2$.
Clearly, when X-ray and SZ data are available for the same object, 
thanks to their different dependence on the electron density,
the joint analysis of both observables allows us to achieve an 
improved description of the thermodynamic properties of the ICM (see e.g., \citealp{2019Eckert}).

In this paper, we focus on the detailed analysis of the protocluster complex 
surrounding the Spiderweb Galaxy (MRC~1138-262; z=2.156), the first galaxy protocluster
for which we have an observational identification of its proto-ICM, provided by both X-ray and SZ data \citep{2022bTozzi,2023DiMascolo}. 
The Spiderweb protocluster was discovered as an overdensity 
surrounding an ultra-steep spectrum radio galaxy \citep{1994Rottgering}
with a mass of $10^{12} \, \rm M_\odot$, as inferred from the K-band luminosity.
This galaxy shows a striking clumpy radio morphology with two jets extending 
from the central source, which shows clear signs of strong interactions 
with the surrounding environment \citep{1997Pentericci}.  
Also, the high rotation measure 
($\sim 6000$ rad/m$^{2}$, discovered in \citealp{1997Carilli}, see also
\citealp{1998Athreya} 
and \citealp{2022Anderson}) confirms the presence of a 
high-density magnetized plasma. Optical observations classified 
the Spiderweb as a narrow emission line galaxy at $z=2.156$ \citep{1997Rottgering}, 
and revealed a clumpy morphology also in the visible band, with a stunning Ly$\alpha$ halo 
extending up to 200 kpc around the central source with a 
luminosity of $\sim 4\times10^{45}$ erg/s (one of the largest known in the Universe).  
Using spectroscopic observations, \citet{2000Pentericci} discovered 15 Ly$\alpha$ emitters within a projected physical distance of 1.5 Mpc from the central radio galaxy. Also, \citet{2004Kurkb, 2004Kurka} found 40 candidate H$\alpha$ emitters. Finally, \citet{2009Hatch} found at least 19 star-forming protocluster members 
with stellar masses in the range of $10^{8}-10^{11}$ $M_{\odot}$ 
within a radius of $\sim$ 150 kpc, which appear to be in the 
process of merging with the central galaxy \citep{2006Miley}. 
Recent X-ray {\sl Chandra} and radio Jansky Very Large Array (JVLA) data allowed 
our team to confirm a clear spatial correlation between the radio 
structures and the Ly$\alpha$ emission \citep{2022Carilli},
providing direct evidence for jet-gas interaction. In addition, 
{\sl Chandra} data revealed an AGN fraction among the protocluster members with log$(M_*/\rm M_\odot)>10.5$
of 25.5$\pm$4.5$\%$ within a radius of $\sim$ 2.5 Mpc, 
one of the highest AGN fraction measured in protoclusters and strongly 
enhanced compared to the field at similar redshift \citep{2022aTozzi, 2023Shimakawa}.
Finally, thermal diffuse emission from a proto-ICM  halo with 
temperature $kT=2.0^{+0.7}_{-0.4}$ keV extending up to a 
radius of 100 kpc has been identified in the region free from the IC emission
associated with the relativistic jets \citep{2022bTozzi}. 

In a study based on data from ALMA \citep{2009Wootten} and the Atacama Compact Array (ACA, a.k.a. Morita Array; \citealt{2009Iguchi}), \citet{2023DiMascolo} 
reported the detection of the SZ effect showing, in an independent and complementary way,
the presence of a nascent ICM within the Spiderweb protocluster. 
The amplitude and morphology of the detected signal reveal that the SZ effect 
from the protocluster is lower than expected from dynamical considerations 
and comparable to that of lower redshift group-scale systems, consistent 
with expectations for a dynamically active progenitor of a local galaxy cluster.
At the same time, this detection clearly shows that current SZ facilities 
could be used to effectively open a novel observational window onto protocluster 
environments. As discussed in \citet{2023DiMascolo}, the combination 
of multi-wavelength data has provided hints for the proto-ICM to be experiencing 
mergers and dynamical interactions with the large-scale protocluster structure, 
the Spiderweb galaxy and its extended jets, and, in general, its multi-phase environment.

In this work we present the combined X-ray and SZ analysis of the proto-ICM of
the Spiderweb protocluster, providing the first spatially resolved characterization of 
an ICM halo at $z>2$. 
The paper is organized as follows.
In Section \ref{sec:data}, we briefly describe the data acquisition and reduction.
In Section \ref{sec:morphology}, we present a detailed analysis of the morphology of the X-ray surface brightness. In Section \ref{sec:combined}, we describe how we perform the SZ and X-ray data combined analysis.
In Section \ref{sec:thermodynamics}, we derive the density, 
temperature and pressure, entropy and cooling time, and the total and ICM mass profiles
of the ICM. 
In Section \ref{sec:discussion}, we provide an estimate of the energy budget associated
with the ongoing feedback processes, the possible presence of a central cooling flow, 
{ the possible implications for the high baryon fraction, and, finally, 
a comparison with other known protoclusters.}
In Section \ref{sec:Future}, we comment on the possible 
extension of this study with SZ facilities from the ground and future X-ray facilities,
such as AXIS and STAR-X.  Finally, our conclusions are summarized in Section \ref{sec:Conclusions}.
Throughout this paper, we adopt the current Planck cosmology
with $\Omega_m =0.315$, and $H_0 = 67.4$ km s$^{-1}$ 
Mpc$^{-1}$ in a flat geometry  \citep{2020Planck}. In this cosmology, at $z=2.156$, 1 arcsec 
corresponds to 8.498  kpc, the Universe is 3.04 Gyr old, 
and the lookback time is 78\% of the age of the Universe.  
Quoted error 
bars correspond to a 1$\sigma$ confidence level, unless noted otherwise.

\section{Observations and data reduction}
\label{sec:data}

\subsection{Chandra X-ray data}
\label{subsec:Chandra X-ray data}

The Spiderweb Galaxy was observed with a {\sl Chandra} Large Program 
of 700 ks with ACIS-S (PI: P.\ Tozzi). The data set includes 21 separate 
observations obtained in November 2019
through August 2020, plus  the first X-ray observation with ACIS-S, 
dating back to June 2000, for additional 39.5 ks \citep{2002Carilli}.
We briefly recall that we run the task 
{\tt acis\_process\_events} with the parameter {\tt apply\_cti=yes} 
to flag background events, most likely associated with cosmic 
rays; by rejecting them, we were able to obtain a significant reduction of 
the background thanks to the VFAINT mode of data acquisition. 
Thus, we exploited the properties of the VFAINT, ACIS-S data
to minimize the noise when deriving the faint surface brightness
of the proto-ICM. The final total exposure time after data reduction and 
excluding the dead-time correction amounts to 715 ks.  
Full details on observations and data reduction procedure done with {\tt CALDB 4.9.3}
can be found in \cite{2022bTozzi}.
The 22 level-2 event files are eventually merged together with the tool {\tt reproject\_obs}, 
using the reference coordinates of Obsid 21483.  
Since this work is focused only on the proto-ICM emission, 
we used the images where the central AGN had been removed after a careful 
rendition of the unresolved AGN, as described in \citet{2022bTozzi}. 
Due to the large luminosity of the AGN, the surface brightness of the
diffuse component within a radius of 2 arcsec is not observationally 
constrained and was treated as a free parameter.   
We used the parameter $n_d$, defined as the ratio of the average 
surface brightness within a radius of 2 arcsec to the average value measured 
between 2 and 3.5 arcsec. {We used this parameter to quantify our lack of information regarding the diffuse emission within 2 arcsec from the nucleus. Considering that we expect the central surface brightness to be larger than in the outer regions, we have $n_d \geq$ 1. We adopted values in the range of 1-6, where $n_d$=1 corresponds to a constant surface brightness; namely,  $n_d \sim 4$ to a developed cool core and $n_d \sim 6$ to an extreme cool core.}
The soft (0.5-2 keV) and hard (2-7 keV) band images in the 
immediate surroundings ($90\times 70$ arcsec$^2$) 
of the Spiderweb Galaxy after the AGN subtraction are shown in Figure \ref{Spiderweb Circular Regions}, 
where we adopted the value $n_d=4$ {(prominent, but plausibly cool core)} as the preferred parametrization of the 
central surface brightness \citep[see][]{2022bTozzi}. 
{ We do not deconvolve the {\sl Chandra} PSF when computing the surface
brightness profile, since the half energy width is $\sim 0.5$ arcsec,
significantly smaller than the minimum width of the annuli we consider here
(2 arcsec). In addition, the surface brightness beyond the central
region (with a radius of 2 arcsec) is smooth, so that the effect
associated with the deconvolution is negligible.}

\subsection{ALMA+ACA data}
\label{sec:alma}
The present ALMA and ACA measurement sets comprise multi-band and multi-configuration 
data, providing a detailed view of the millimeter footprint of the Spiderweb galaxy 
and its surroundings. In particular, we considered targeted Band 3 
($94.5-110.5~\mathrm{GHz}$) observations obtained by ACA and ALMA 
(project code: 2018.1.01526.S, PI: A. Saro), spanning $uv$ ranges 
of $2.2-17.5~\mathrm{k\lambda}$ and $3.93-1320~\mathrm{k\lambda}$, respectively (corresponding to $767-97~\mathrm{kpc}$ and $434-1~\mathrm{kpc}$ 
at the redshift of the Spiderweb protocluster). To better characterize the 
spectral properties of the Spiderweb galaxy and exploit the peculiar spectral 
scaling of the SZ effect for facilitating its separation from the extended 
radio emission, we complemented the observations with archival 
Band~4 observations with ALMA (project code: 2015.1.00851.S, PI: B. Emonts) 
and ACA (project code 2016.2.00048.S, PI: B. Emonts). 
The data calibration and processing procedures, described in full detail 
in \citet{2023DiMascolo} and consisting a standard pipeline calibration applied to 
all the Band 3 ALMA and ACA data using the Common Astronomy Software 
Application (CASA\footnote{\url{https://casa.nrao.edu/}}). 
Instead, we obtained the calibrated Band 4 measurement sets through the 
calMS service \citep{2020Petry} offered by the European ALMA Regional 
Centre network \citep{2015Hatziminaoglou}. The direct inspection of all 
the available data sets did not highlight any issue with the standard calibration, thus, we did not perform any additional calibration, flagging, 
or post-processing steps.

Finally, we note that in this work, we are interested in inferring 
solely the large-scale morphology and thermodynamic distribution of 
the proto-ICM in the Spiderweb protocluster. As such, we considered 
only the subset of observations below a threshold in $uv$ distance of 
$65~\mathrm{k\lambda}$. As detailed in \citet{2023DiMascolo}, such a 
choice makes the extended radio galaxy unresolved along the direction 
perpendicular to the jet direction. This drastically reduces the morphological 
components required to model the emission, in turn, simplifying the overall 
modeling procedure. All the SZ results reported hereafter 
refer only to the specific $<65~\mathrm{k\lambda}$ subset of ALMA+ACA measurements.

\section{Morphology of X-ray emission}
\label{sec:morphology}

\begin{figure*}
   \centering
   \includegraphics[scale=0.3]{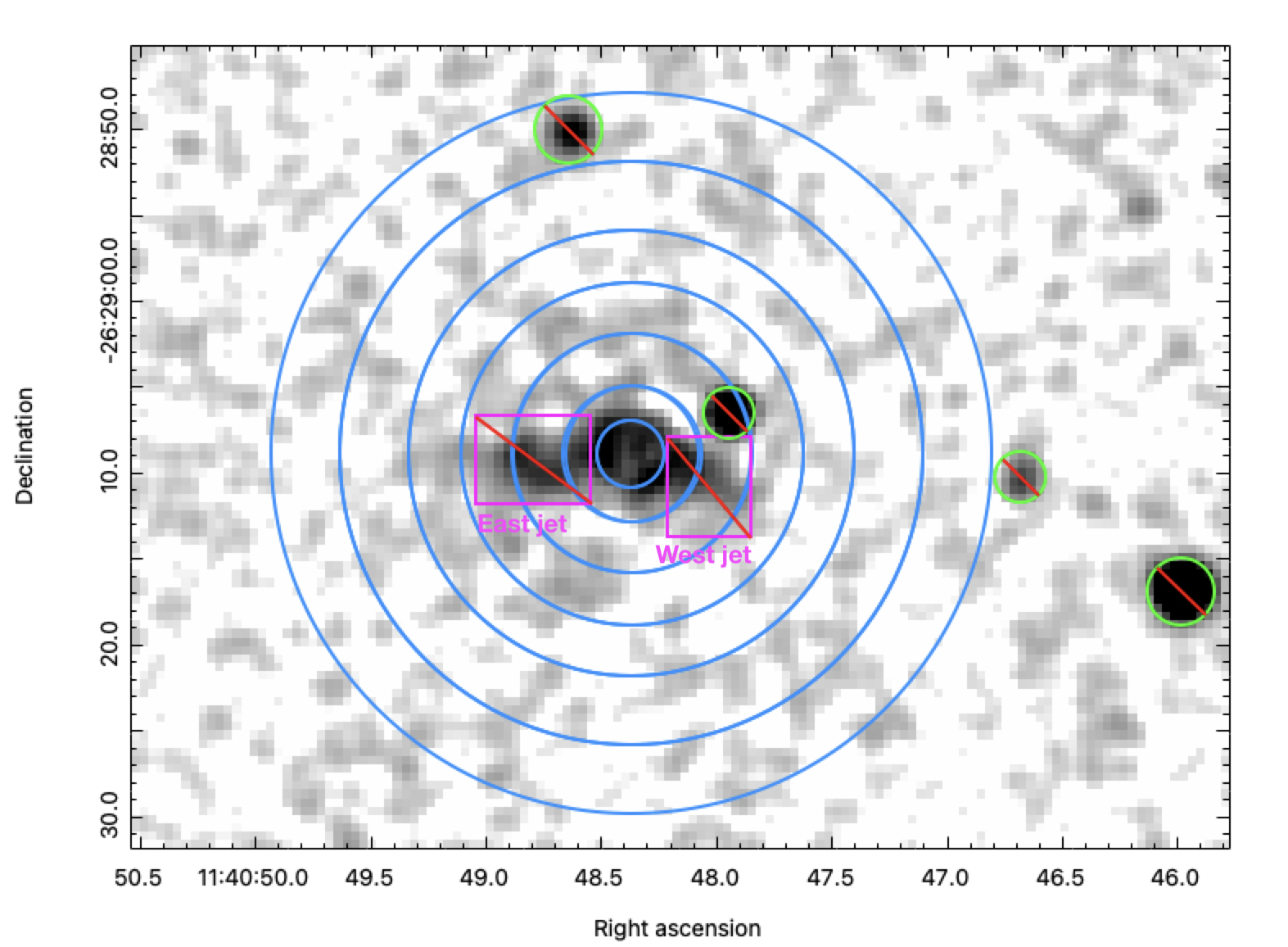}
   \includegraphics[scale=0.3]{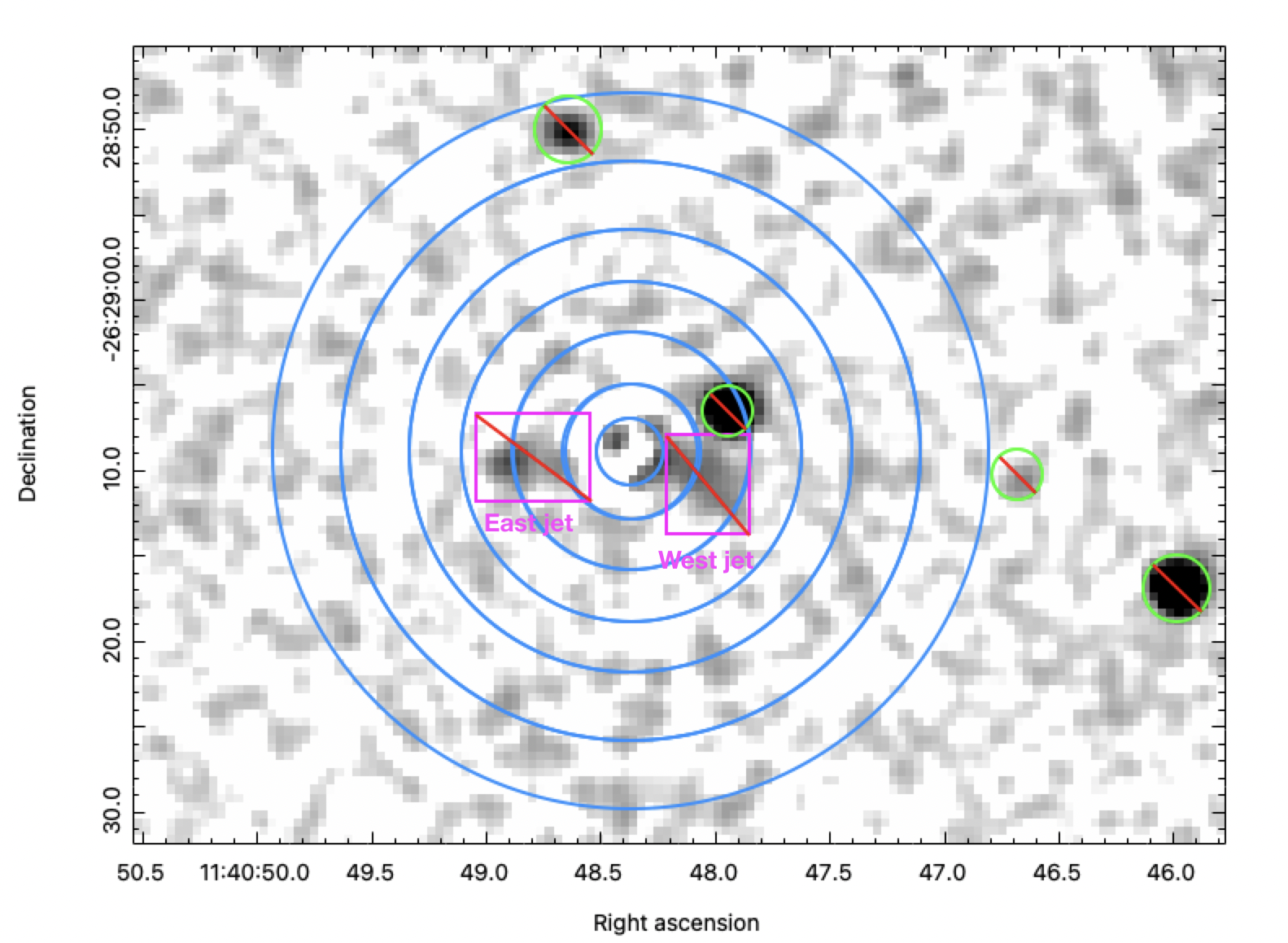}
   \includegraphics[scale=0.5]{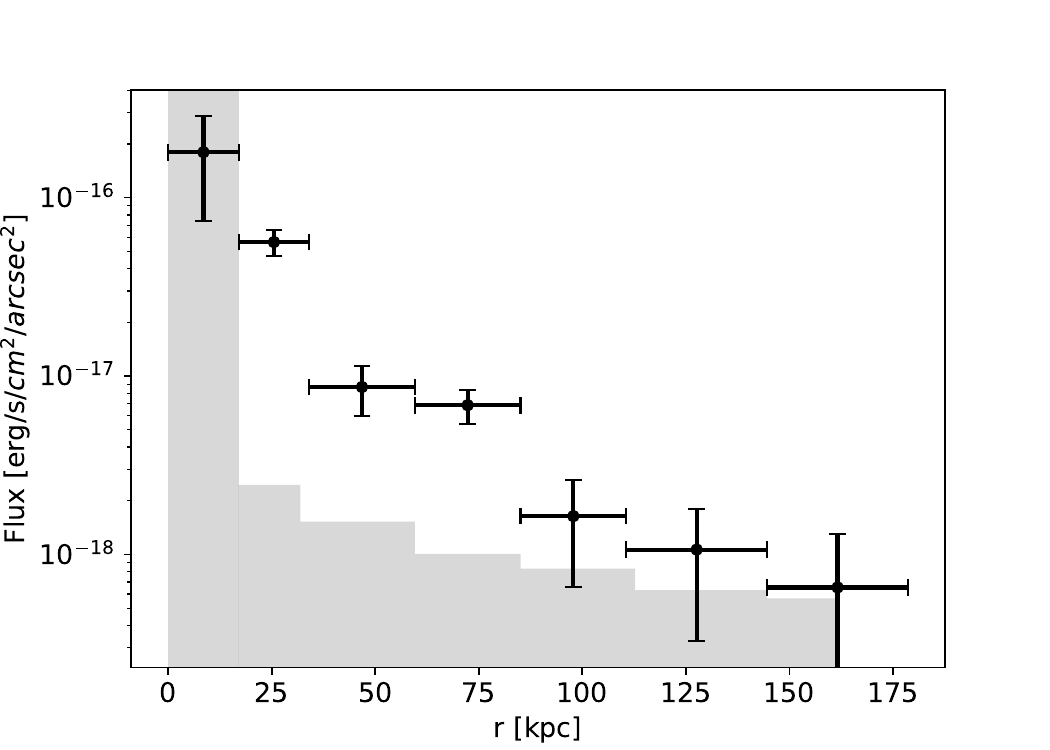}
   \includegraphics[scale=0.5]{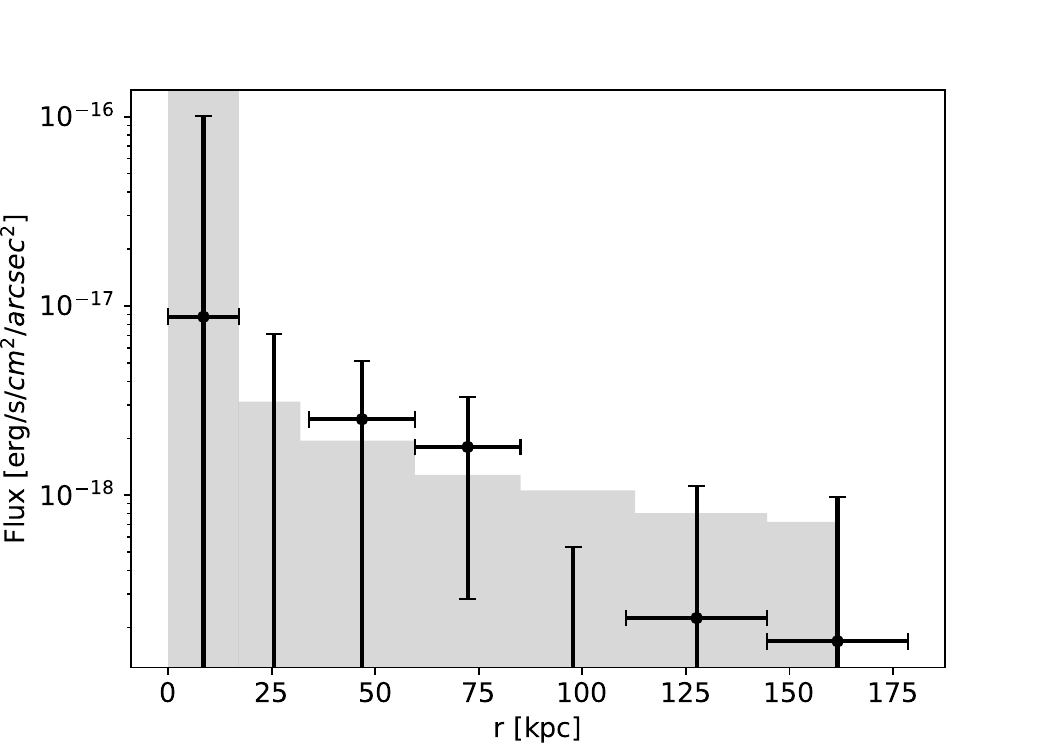}
   \caption{AGN-subtracted soft (0.5-2 keV, upper left panel) 
   and hard (2-7 keV, upper right panel) band images of the Spiderweb protocluster in a field of view of $\sim$ $90\times 70$ arcsec$^2$.
   The blue concentric circles represent the regions within which we derived 
   the surface brightness and correspond to an outer radius of 2, 4, 7, 10, 13, 17, and 21 arcsec, 
   with the central blue circle showing the region where the unresolved AGN emission makes it impossible to estimate the diffuse emission. The green circles represent removed unresolved sources and the magenta boxes represent the eastern and western jet regions overlapping with the radio emission, and are excluded from our analysis since they are dominated by the diffuse, non-thermal IC emission from the jets. For the purposes of visualization, the central surface brightness is parametrized with $n_d=4$, as discussed in \citet{2022bTozzi}.  In the lower-left and lower-right panels, we show the 
   corresponding background-subtracted surface brightness profiles computed 
   according to Equation \ref{sb_equation}. 
   The black bars represent the 1$\sigma$ uncertainties associated with data points 
   including the Poissonian errors on the total counts and the background 
   expected counts. The grey-shaded area marks the value of surface brightness parametrized with $n_d=4$ in the first bin and the 1$\sigma$ background level in the other bins.}
    \label{Spiderweb Circular Regions}
\end{figure*}

{As stated in Section \ref{sec:Introduction}, in
the region free from the jets emission, \cite{2022bTozzi} 
ascribed the thermal diffuse X-ray emission to the presence of 
a proto-ICM halo extending up to a radius of 100 kpc. Indeed,
from a visual inspection of the X-ray images,
we can immediately identify the diffuse, spherically symmetric emission that is contributed 
by the proto-ICM in the X-ray soft band (upper left panel of Figure \ref{Spiderweb Circular Regions}), 
while diffuse emission in the hard band is observed only in correspondence with 
radio jets; therefore, the latter is dominated by the non-thermal IC emission.  
As a consequence, as discussed in \citet{2022bTozzi}, we have only a partial view of 
the surface brightness distribution of the proto-ICM, since in the two regions 
overlapping with the radio jets, the IC emission originated by the relativistic electrons
is larger than the thermal emission in the soft band.} 

{Therefore, in this work we assume that all
the diffuse emission outside the jet and AGN regions is associated with the
proto-ICM.  Nevertheless, we have not been able to firmly exclude a
non-thermal contribution.  For example, in the
case of the radio galaxy 3C294, the extended emission perpendicular to
the jets is tentatively associated with IC from an old population of
relativistic electrons spread over a large region, due to the precession
of the jets on a timescale of 100 Myr \citep{2003Fabian,2006Erlund}.
However, in the Spiderweb galaxy we do no find any hint of jet
precession, when, in fact, it should be visible since it would occur on a
timescale comparable to the age of the jet \citep{2022Carilli}. In
addition, the X-ray surface brightness of the Spiderweb is softer than
a typical $\Gamma \sim 2$ power law, and it does not show sharp edges as
in the case of 3C294, which is a feature that may be associated with a population of
pressure-confined relativistic electrons.  Needless to say, we are not
able to explore the presence of low surface brightness, roughly
isotropic radio emission in our L band JVLA or Giant Metrewave Radio Telescope (GMRT) data, due to the
limited dynamic range caused by the presence of the bright radio
galaxy. Overall, we recognize that we cannot definitively exclude an IC
contribution to the X-ray diffuse emission; however, this scenario is not
supported by any observed feature.  We go on to discuss the effects
of a possible non-thermal contribution in the discussion in Section 6.}

{Thus, considering the soft, 
diffuse X-ray emission as entirely associated with the presence of a 
proto-ICM,} we are able to detect such emission only 
beyond 2 arcsec due to the very high AGN luminosity, while the X-ray luminosity associated 
with the proto-ICM within 2 arcsec can  only be parametrized by assuming a reasonable 
electron density profile. Clearly, all
the other unresolved sources identified in the field are also masked out.

To explore the properties of the X-ray surface brightness associated with the proto-ICM 
thermal emission, we proceeded according to the following steps:\ 1) we calculate the number of total counts within specific selected regions 
    in the soft and hard bands, excluding unresolved sources and jet regions; 2) we calculate the area of each selected region after removing the masked areas; 3) we sample the soft and hard band background from a source-free annulus with inner
    and outer radius of 16 and 29.5 arcsec, respectively, as described in \citet{2022bTozzi}; 4) we calculate the vignetting correction in the soft and hard bands, using the monochromatic
    exposure maps at 1.5 and 4.5 keV, respectively; 5) we then compute the net detected counts and associated Poissonian 
    uncertainty after subtracting the 
    background rescaled by the geometrical area and applying the vignetting correction; 6) using XSPEC v.12.12.0, we calculate the conversion factor from the count rate to energy flux, assuming a thermal spectrum with $kT=2$ keV and a Galactic 
    absorption of $N_{H}^{Gal}=3.18 \times 10^{20}$ cm$^{-2}$ according to the HI map of
    the Milky Way \citep{2016HI4PI}; 7) finally, we obtain the average surface brightness of each selected region in units of erg s$^{-1}$ cm$^{-2}$ arcsec$^{-2}$. These steps are synthesized by the relation: 
    
    \begin{equation}
        SB=ECF \times \frac{(CTS - BCK \times A_S/A_{BCK})}{T_{exp}\times A_S}
        \times \frac{Expmap_{Max}}{Expmap_S}
        \label{sb_equation}
    ,\end{equation}
    where \emph{ECF} is the energy conversion factor from count rate to energy flux and is equal to $\sim 1.20\times10^{-11}$ erg/s/cm$^{2}$/cts in the soft band and $\sim 1.24\times10^{-11}$ erg/s/cm$^{2}$/cts in the hard band; 
    \emph{CTS} is the number of total counts observed in the image within a given region; 
        \emph{BCK} is the number of total counts observed in the background region; \emph{$A_S$} and 
        \emph{$A_{BCK}$} are the area values corresponding to the selected 
        region and the background region, respectively; { $T_{exp}$ is the total exposure time of the observation after removing the dead time intervals and possible high background rate intervals. 
        The ratio \emph{$Expmap_{Max}$/$Expmap_S$} is the 
        vignetting correction that is applied to the net counts
        (observed where the exposure map has the value $Expmap_S$)
        to compute the signal that would be obtained at the aimpoint (where the 
        exposure map reaches the maximum value $Expmap_{Max}$). The maps are computed separately in the soft and hard bands and describe the effective area in $cm^{2}$ at each position in the field of view and, thus, bring the information on the sensitivity of the instruments. }
        Since the signal from the surface brightness is very low, our approach consist in 
        measuring the surface brightness in regions selected specifically to identify 
        possible features of the proto-ICM projected distribution and quantify the 
        corresponding statistical significance.

\subsection{Azimuthally averaged profile}
\label{sec:Azimuthally-averaged profile}

The first step is to derive the surface brightness profile, 
assuming a spherically symmetric distribution centered on the 
AGN position (RA=11:40:48.3611 and Dec=-26:29:08.984).  
Therefore, we extracted the surface brightness profile from 
seven annuli with outer radius of 2, 4, 7, 10, 13, 17, and 21 arcsec
centered on the AGN for the soft and hard images, as shown in the upper panels of 
Figure \ref{Spiderweb Circular Regions}.
Applying Equation \ref{sb_equation}, we obtain the surface brightness profile in 
erg/s/cm$^{2}$/arcsec$^{2}$ in the soft and hard bands, as shown in the lower panels of 
Figure \ref{Spiderweb Circular Regions}. We find that 
the azimuthally average surface brightness is well detected in the soft band 
at radii $r<150$ kpc ($\sim 17$ arcsec), while in the hard band, the photometry 
is consistent with no signal in all the annuli. These findings confirm that the 
diffuse emission is entirely soft, consistent with being bremsstrahlung
emission with an estimated average temperature of $\sim 2$ keV, as derived
from the X-ray spectral analysis in \citet{2022bTozzi}.  
In the rest of the paper, we work on the soft band image only and we also neglect the data beyond 17 arcsec in the soft band, 
due to the drastic drop in the number of net counts which could be associated with the ICM.

We also note that the surface brightness profile is strongly increasing towards the center. 
As said, we are not able to measure the diffuse emission in
the innermost region included within a radius of 2 arcsec from the
nucleus because of the overwhelming
AGN X-ray brightness.  As discussed in \citet{2022bTozzi}, we assume a reasonable estimate
for the central brightness of the thermal emission corresponding to 
a concentration value of $n_d=4$. Therefore, the innermost region where we are
actually able to measure the ICM emission is the annulus between 2 and 3 arcsec
(corresponding to distances of $\sim$16 and $\sim 25$ kpc, respectively). 
Thus, as shown in Figure \ref{Spiderweb Circular Regions}, we observe an increase of a factor of five
from 50 kpc to $\sim 25$ kpc in the soft-band surface brightness. 
Also, we notice that the subtraction of the central AGN does not introduce
systematics, since the same subtraction procedure results in a null profile in the 
hard-band, suggesting that the AGN subtraction left no significant residuals.
Therefore, we concluded that the soft band image provides the first strong hint 
for a significant cool core in the Spiderweb Galaxy on a scale of $\sim 30$ kpc.

\begin{figure*}
   \centering
   \includegraphics[scale=0.26]{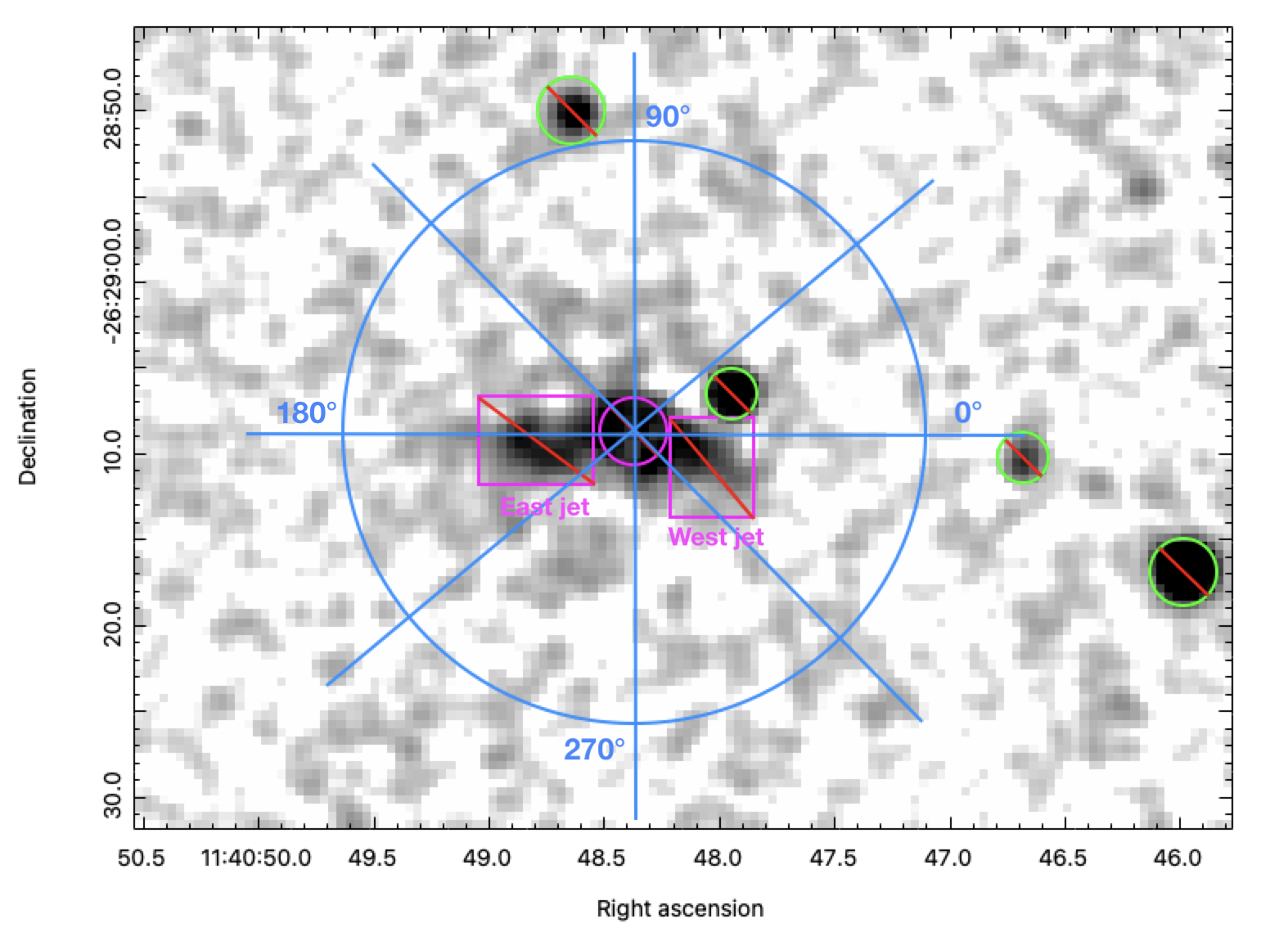}
      \includegraphics[scale=0.5]{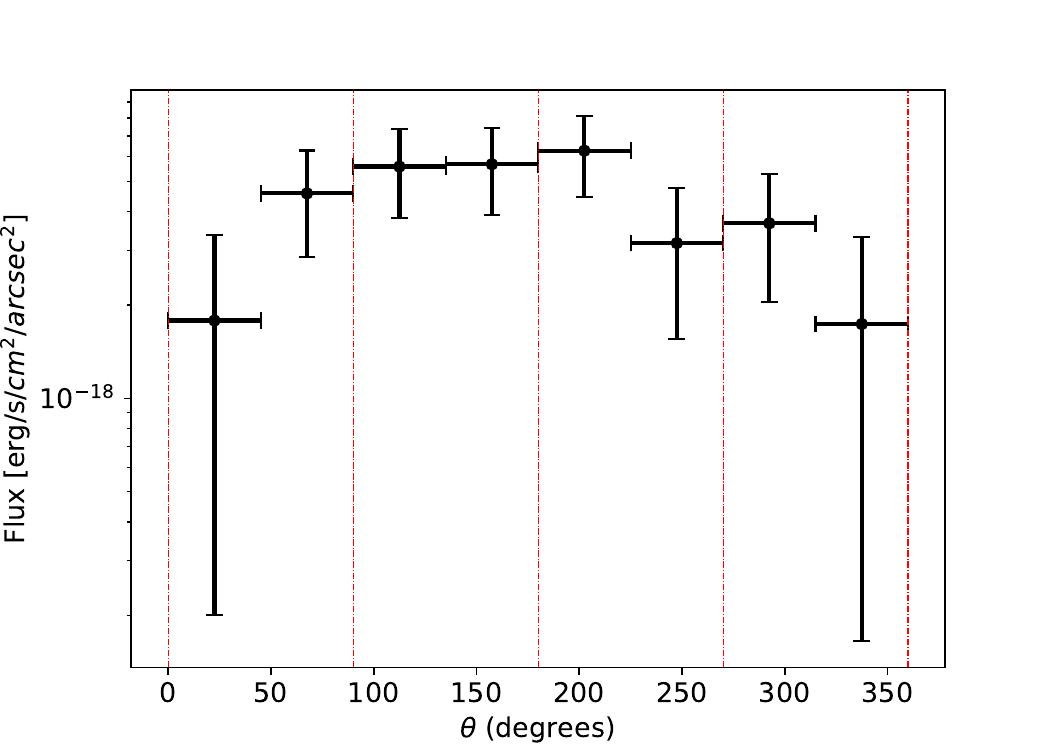}
   \caption{AGN-subtracted soft band image of the Spiderweb 
   protocluster (left). The blue circle corresponds to 17 arcsec, while the blue lines 
   represent the regions within which we derived the surface brightness 
   as a function of the angle. 
   Magenta squares and magenta circle represent the eastern and western 
   jet excluded regions and the central excluded AGN, respectively, while the green 
   circles represent the excluded unresolved sources in the field of view. 
   The average surface brightness within each wedge in the soft band is presented as a function 
   of the angle (right). The black bars represent the Poissonian uncertainties associated 
   with surface brightness measurements. The red dotted lines separate different quadrants.}
    \label{wedges_1}
\end{figure*}

\subsection{Azimuthal fluctuations}
\label{sec:Azimuthal fluctuations}

The previous measurement of the surface brightness distribution relies on two assumptions that are
usually reasonable in the context of  virialized massive halos: a center that is coincident with the central galaxy
and a spherically symmetric distribution. Here, instead, we explore the presence of azimuthal 
fluctuations. To do so, first we considered eight different wedges with a step of 45 degrees (as shown in the left panel of Figure \ref{wedges_1}), extending up to a distance of 17 
arcsec from the central AGN and excluding the central 2 arcsec and the jet regions.
The average surface brightness in each wedge as a function of the position angle is shown in 
the right panel of Figure \ref{wedges_1}.  We find that the azimuthal fluctuations 
of the average surface brightness in the wedges are consistent with  isotropy. If we fit
the points shown in the right panel of Figure \ref{wedges_1} with a 
constant surface brightness as a function of the
azimuthal angle, we find a reduced value, $\chi^{2}=1.1,$ with 7 d.o.f.
To explore the effect of the wedge width, we 
repeated the measurement using 12 wedges with a step of 30 degrees.
In this case, we obtained a reduced $\chi^{2}=1.18$ for 11 d.o.f. by assuming a constant 
surface brightness as a function of the azimuthal angle.
This slight increase of the reduced $\chi^2$ is mostly due to a depression visible at $\sim 250$ degrees. This may point towards 
an actual decrease in the ICM surface brightness in this sector.

\begin{figure*}
   \centering
   \includegraphics[scale=0.25]{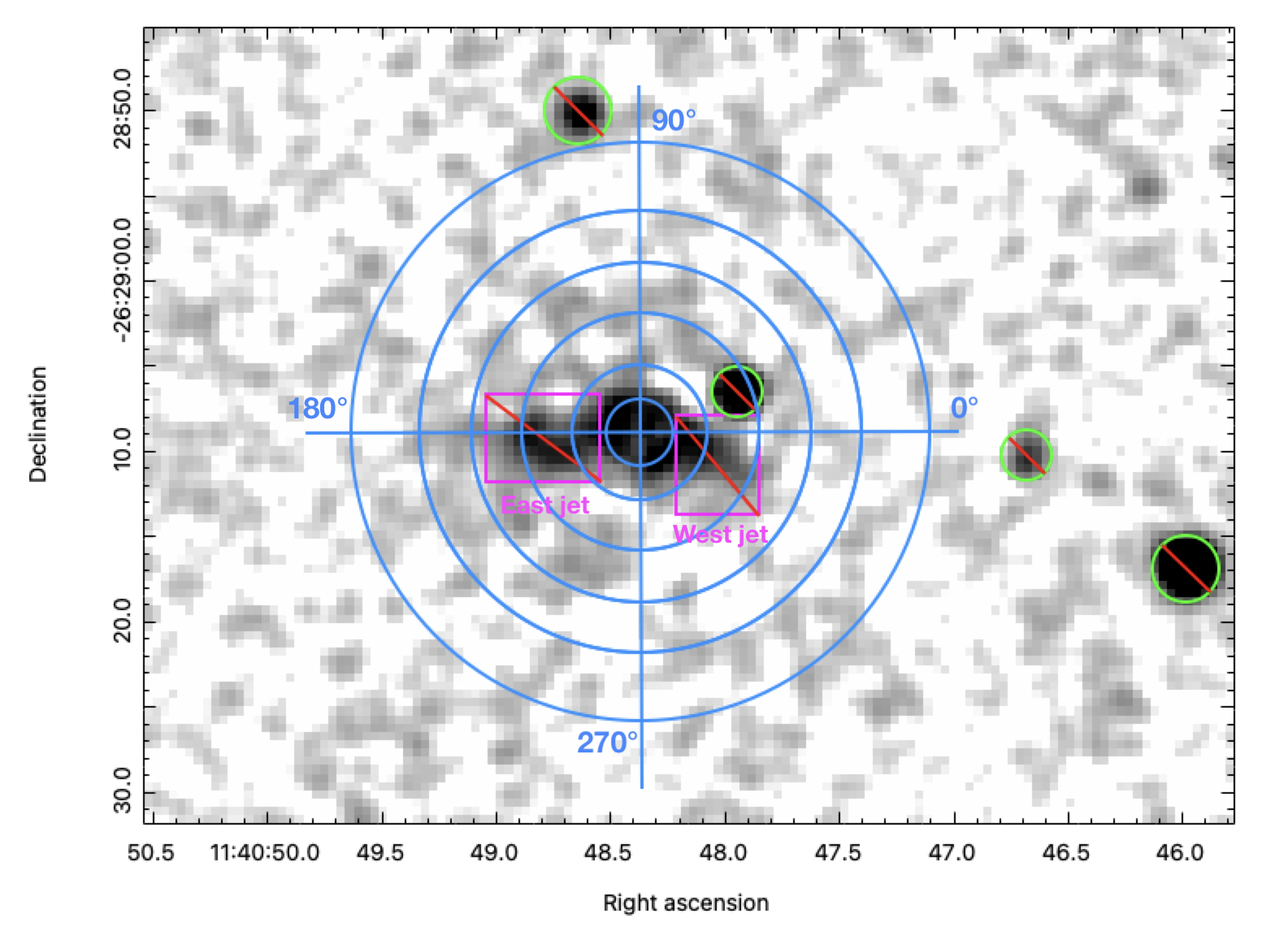}
   \includegraphics[scale=0.5]{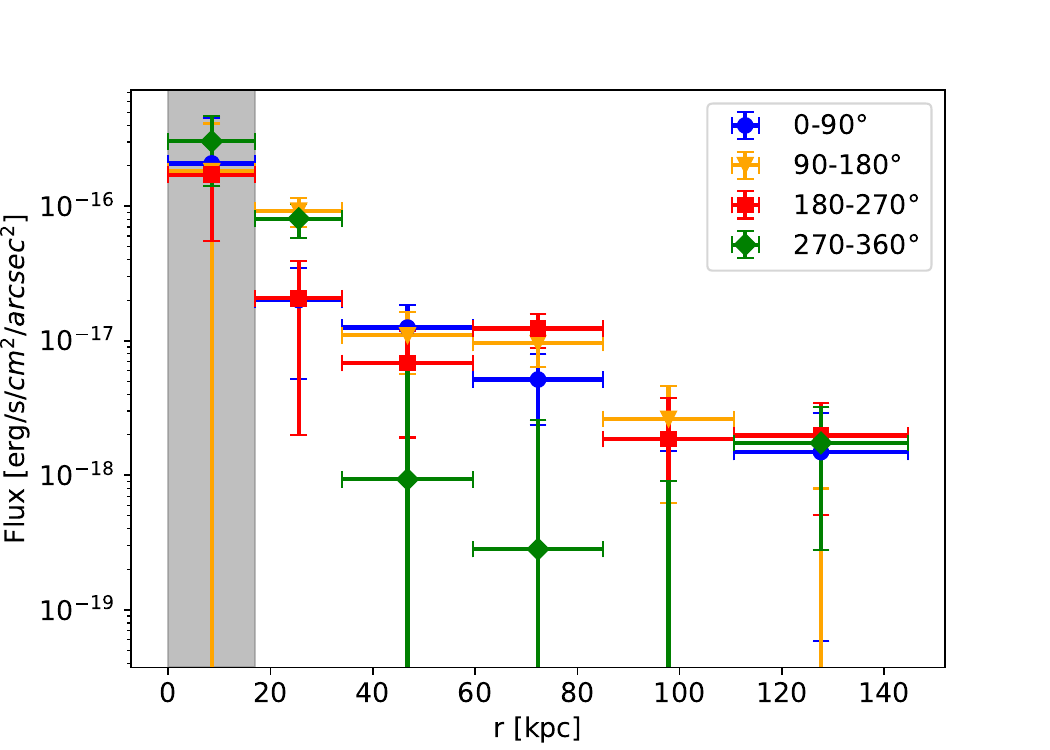}
    \caption{AGN-subtracted soft band image of the 
    Spiderweb protocluster (left). The blue concentric circles, which correspond 
    to 2, 4, 7, 10, 13, and 17 arcsec, and the blue lines separate the quadrants within which we derived the surface brightness as a function of the angle. 
    Magenta squares represent the eastern and western 
   jet excluded regions, while the green 
   circles represent the excluded unresolved sources in the field of view.  Surface-brightness profiles
    as a function of the distance from the central radio source (right). The grey-shaded area represents the surface brightness values parametrized with $n_{d}$.}
    \label{Spiderweb Slices + Annuli}
\end{figure*}

To further investigate the presence of azimuthal 
changes in the surface brightness distribution, 
we considered four quadrants (0-90, 90-180. 180-270, 270-360 degrees) 
and extracted the surface-brightness profile from circular concentric 
regions at a distance of 
2, 4, 7, 10, 13, and 17 arcsec from the central source 
(see Figure \ref{Spiderweb Slices + Annuli}, left panel). 
The four surface brightness profiles 
obtained are shown in Figure \ref{Spiderweb Slices + Annuli} (right panel), where 
a different color {and symbol} is associated with a different quadrant. 
We see a drop in the south-west quadrant in at least two bins, 
but with a statistical significance lower than 2 $\sigma$. 

Overall, we conclude that the morphology of the X-ray emission can be 
reasonably approximated with a spherically symmetric surface brightness 
distribution centered on the AGN.  Nevertheless, we also find hints of 
a departure from symmetry in the form of a depression of about 
an order of magnitude 
in the same quadrant where we find the western jet.  Clearly, it is not 
possible to associate this depression
to the effect of the jet, however, we argue that this feature 
may be associated with a cavity carved by the jet into the proto-ICM. 
Alternatively, this can also be a feature associated with the
disturbed dynamics of the halo, or with the chaotic
turbulent eddies which generate both compressions and depressions in
the ICM \citep[see X-ray maps in][]{2017Gaspari}.  { To further test
these results, we explore the morphology of the SZ signal in the next section.}

\section{SZ and X-ray combined analysis of the proto-ICM distribution}
\label{sec:combined}

In the sections above, we describe how the X-ray surface brightness distribution 
of the proto-ICM  in the Spiderweb protocluster reveals a depression 
in the south-west quadrant. Therefore, we would expect to find that the 
centroid of the X-ray emission is not centered on the radio galaxy, but
is shifted towards the east. 
To quantify this shift, we analyzed the \emph{Chandra} X-ray 
images with the \emph{Sherpa} software in {\tt ciao}.

\begin{figure}
   \centering
   \includegraphics[width=0.53\textwidth]{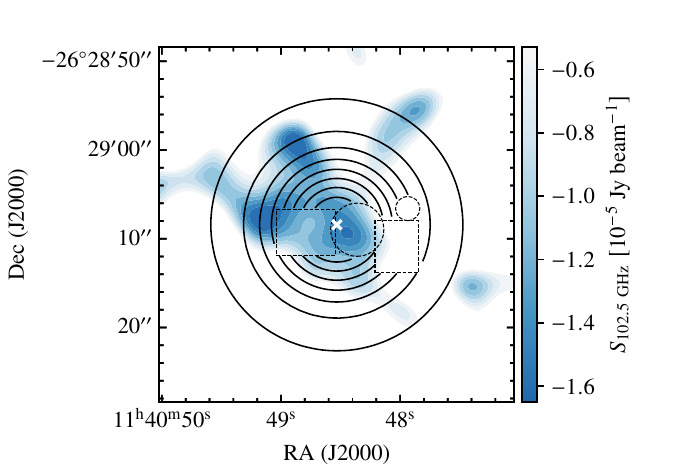}
    \caption{Adapted from \cite{2023DiMascolo}: Contours from the SZ surface brightness map from a combined ALMA+ACA image (blue). The contours are set arbitrarily to optimally emphasise the bulk SZ signal. The black solid contours represent the model obtained with the single two-dimensional beta model fitting on \emph{Chandra} AGN-subtracted X-ray image in the soft band, while the dotted circles and dotted squares represent the regions excluded in the analysis, namely, the eastern and western jets, central AGN, and an unresolved source. The white cross represents the X-ray emission centroid.}
    \label{X-SZ}
\end{figure}

We considered the AGN-subtracted \emph{Chandra} X-ray image in the soft band. 
However, here we mask a larger portion of the central region, removing a circle of 3 
arcsec centered on the AGN.  In this way, we removed 
the spike in the surface brightness distribution
that we tentatively identified as a cool core \citep[see][]{2001MolendiPizzolato}, 
which is further analyzed in 
the  sections below.  Then, we fit the surface brightness distribution 
with a single two-dimensional (2D) beta model plus background 
in order to find the X-ray centroid of the emission.
As we can see from Figure \ref{X-SZ}, the X-ray emission centroid (white cross) 
is shifted towards the east by $\sim 2.3$ arcsec ($\sim 20$ kpc) with an uncertainty of $\sim 0.6$ arcsec. 

Consistently with the findings of \citet{2023DiMascolo}, the pressure distribution of the proto-ICM 
electrons was instead reconstructed by means of a Bayesian forward 
modeling of the raw $uv$-plane interferometric data from Band~3 and 
4 ALMA+ACA observations (Section \ref{sec:alma}). Briefly, the SZ distribution 
is obtained as the line-of-sight integral of a generalized Navarro-Frenk-White (NFW, \citealp{2007Nagai}) pressure distribution, with scaling parameters and radial slopes 
fixed to match specific formulations from the literature. In particular, 
we considered the full set of pressure models employed in \citet{2023DiMascolo}. 
An overview of the pressure model details can be found in Appendix~\ref{app:sz}.
For consistency with the X-ray analysis discussed above, we did, however, fix the 
centroid position of the SZ model component to match the one derived 
by modeling the soft-band X-ray surface brightness distribution. 
To fully account for the cross-contamination between the SZ signal and 
the extended emission from the Spiderweb radio galaxy, we inferred the 
SZ component jointly with the radio galaxy model, defined by the optimal, 
ordered set of point-like components, as in \citet{2023DiMascolo}. 
We also consistently imposed an upper threshold of $65~\mathrm{k\lambda}$ 
to the analysed visibilities to make the extended radio galaxy 
unresolved in the direction orthogonal to the jet axis. The posterior exploration 
was performed using the nested sampling algorithm \citep{2004Skilling,2022Ashton}, 
as implemented in the \texttt{dynesty} package \citep{dynesty,2022Koposov}. 

\begin{figure}
   \centering
   \includegraphics[scale=0.523]{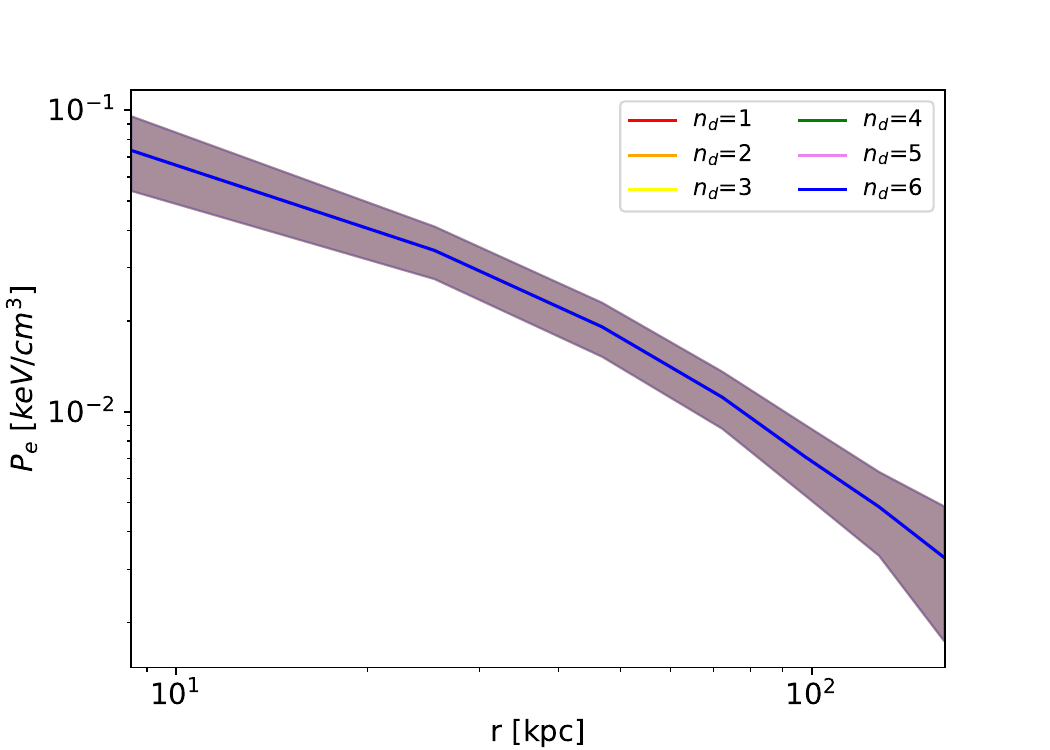}
   \caption{Best-fit pressure profile 
   obtained from SZ signal by marginalizing over the different 
   gNFW models using a BMA technique. The solid lines denotes 
   the median pressure profile, while the shaded regions mark 
   the 68\% credible interval on the posterior BMA model. 
   We use different colors for different $n_{d}$, but we can see 
   the pressure profile is not affected by this parametrization. 
   }
   \label{fig:pressure_profile}
\end{figure}

Despite the centroid assumption introduced above, the parameters 
for each of the reconstructed models are found to be statistically consistent 
with the ones reported in \citet{2023DiMascolo}, whereby the SZ shift with respect to the central AGN is 6.2 $\pm$ 1.3 arcsec. In this regard, we note that 
we observe a systematic reduction of the mass estimates, but they fall well 
within the respective statistical uncertainties, providing support to the 
robustness of the derived mass parameters (for details on the inferred models we refer to Appendix~\ref{app:sz}, the corresponding Table~\ref{app:tab:sz} and also to \citealp{2023DiMascolo}). 
In Figure \ref{fig:pressure_profile} 
we show the pressure model obtained by means of 
Bayesian Model Averaging (BMA; \citealt{1999Hoeting}, \citealt{2018Fragoso})
applied to the different flavours of pressure profiles inferred 
with our analysis. Hereafter, we refer to the BMA model as the 
reference profile for the present study and we propagate the variance 
between the various pressure models in the marginalized posterior distribution.

The low significance of both the SZ signal and the X-ray emission 
from the proto-ICM inevitably limits the possibility of performing 
a detailed morphological comparison. However, as already hinted at 
in \citet{2023DiMascolo}, we observe a marginal agreement between the 
bulk distribution of the proto-ICM inferred from the SZ signal and from 
the X-ray diffuse emission beyond 3 arcsec from the AGN (see Figure \ref{X-SZ}).  This result provides 
a robust and straightforward confirmation of the global morphological 
picture obtained in both observational windows,
confirming a substantial asymmetry consisting in a shift 
of the bulk of the ICM toward the east with respect to the Spiderweb galaxy.
Alternatively, we may have identified a depression (tentatively a cavity, given the directional correspondence with the western radio jet) 
in the south-west quadrant. 
Unfortunately, we are not able to provide further constraints on the nature of the 
asymmetry in the proto-ICM with current data. 
As previously mentioned, we may interpret our result on the off-centered distribution of the ICM as 
a sign of the dynamical state remaining far from equilibrium or, alternatively, of ongoing radio-mode feedback. 
Both scenarios are further discussed in Section \ref{sec:discussion}. 
Henceforth, we focus on the 
spherical approximation, to derive physical properties of the proto-ICM.

\section{Thermodynamic properties}
\label{sec:thermodynamics}

In this section, we analyze the thermodynamic properties of the proto-ICM in the halo of
the Spiderweb Galaxy.

\begin{table*}
\caption{{Parameters and derived quantities of the beta model fit.} }
\label{table:beta model parameters}      
\centering          
\begin{tabular}{c  c  c  c}     
\hline\hline       
$n_{d}$ & $r_{c}$ & $S_{0}$ & $n_{e0}$ \\
& [arcsec] & [$10^{-17}$$erg/s/cm^2/arcsec^2]$ & [$cm^{-3}]$ \\
\hline                    
    1 & 2.75$\pm$0.18 & 1.48$\pm$0.17& 0.132$\pm$0.011 \\
    2 & 2.58$\pm$0.16 & 1.72$\pm$0.19& 0.143$\pm$0.011\\
    3 & 2.42$\pm$0.14 & 2.02$\pm$0.21& 0.156$\pm$0.012\\
    4 & 2.25$\pm$0.13 & 2.41$\pm$0.25& 0.172$\pm$0.012\\
    5 & 2.09$\pm$0.11 & 2.87$\pm$0.29& 0.189$\pm$0.013\\
    6 & 1.92$\pm$0.10 & 3.57$\pm$0.36& 0.213$\pm$0.015\\
\hline                  
\end{tabular}
\tablefoot{$n_{d}$ parametrizes the diffuse emission inside the central 2 arcsec in which resides the AGN emission, $r_{c}$ is the core radius, $S_{0}$ is the central value of surface brightness, and $n_{e0}$ is the central value of electron density.}
\end{table*}

\subsection{ICM density profile}
\label{subsec:density_profile}

According to the standard $\beta$-model \citep[][]{1966King,1976CavaliereFuscoFermiano} 
the ICM is assumed to be an isothermal 
plasma in equilibrium with the galaxies in the same potential. 
Assuming that the galaxy distribution can be approximated as $\rho_{gal}(r)=\rho_{gal0}[1+(r/r_{c})^{2}]^{-3/2}$ \citep[][]{1978CavaliereFuscoFermiano}, the gas density, 
$n_{gas}$, can be written as:

\begin{equation}
    n_{gas}(r)=n_{gas0}\biggr[1+\biggr(\frac{r}{r_{c}}\biggr)^{2}\biggr]^{-3\beta/2}\, , 
    \label{eq: gas density}
\end{equation}

\noindent
where $r_{c}$ is the core radius and $\beta=\mu m_{p} \sigma_{r}^{2}/kT_{g}$ 
(with $\mu$-mean molecular weight, $m_{p}$-proton mass, $T_{g}$-gas 
temperature and $\sigma_{r}$-one-dimensional velocity dispersion). 
Consistently with Equation \ref{eq: gas density}, the X-ray 
surface brightness profile at a projected core radius, $r,$ is expressed as:

\begin{equation}
    S(r)=S_{0}\biggr[1+\biggr(\frac{r}{r_{c}}\biggr)^{2}\biggr]^{-3\beta+1/2}\, .
    \label{eq: SB profile}
\end{equation}

\noindent

\begin{figure}
\centering
   \includegraphics[scale=0.4]{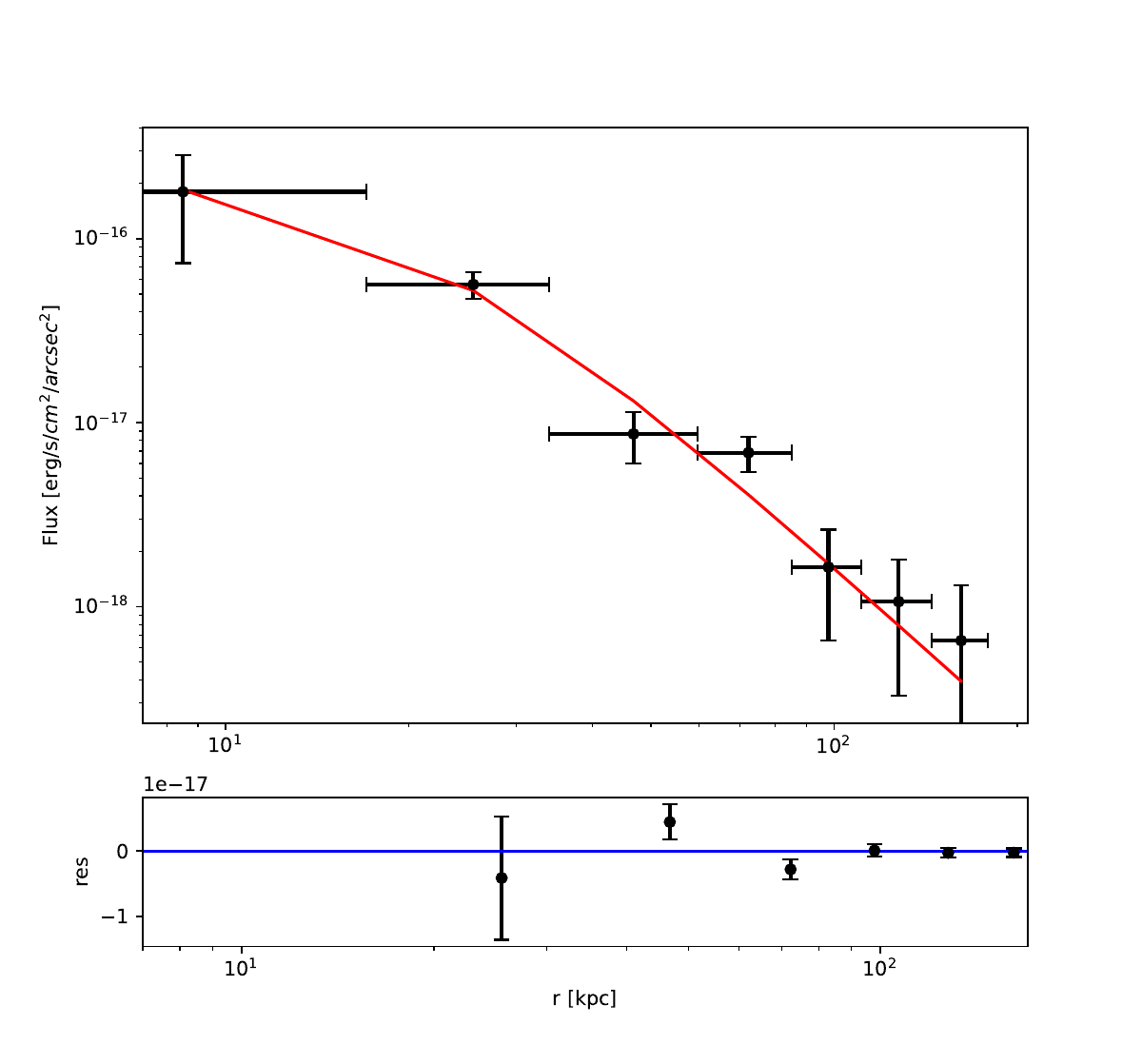}
   \caption{Best-fit 
    of the surface brightness profile in the soft band for $n_{d}$=4 as a function 
    of the distance from the AGN,
    shown as the red solid line, considering the extraction from the annular regions in 
    Figure \ref{Spiderweb Circular Regions} (top panel). The X-ray surface brightness 
    is shown with solid dots and associated error bars. Data points are
    the same as in Figure \ref{Spiderweb Circular Regions}. Fit residuals are given in the bottom panel.}
    \label{Spiderweb Annuli Fit}
\end{figure}

Typically, a value of $\beta\sim0.66$ is a good, physically motivated approximation 
to describe the X-ray surface brightness and density profile in galaxy clusters \citep{1976CavaliereFuscoFermiano, 1978CavaliereFuscoFermiano}.
The $\beta$ model is characterized by three parameters
(normalization, core radius, and external slope). Measurements of the three offer a complete meaningful description 
of a spherically symmetric, self-gravitating halo. However, due to the 
well known degeneracy between $r_c$ and $\beta$, these two parameters may be simultaneously 
constrained only when the surface brightness can be probed over a broad range of radii from $r<<r_c$ to $r>>r_c$.
In our case, given the low signal from the proto-ICM in the Spiderweb protocluster, 
we are limited only to the central regions, at the point where we can approximate
the proto-ICM distribution with a constant density, as we did in \citet{2022bTozzi}.
For the purpose of the analysis in this paper, to characterize the Spiderweb proto-ICM 
we fix $\beta=2/3$.
The center of the $\beta$ model is fixed to the AGN coordinates and, therefore, it 
does not imply any additional free parameters.  
First, we fit the surface brightness profile,  
obtaining the values of $S_0$ and $r_{c}$ for different $n_{d}$ 
values reported in Table \ref{table:beta model parameters}.  
The best-fit model for $n_{d}=4$ is shown in Figure \ref{Spiderweb Annuli Fit}. 
{ Error bars (at 1$\sigma$ confidence level) 
on the best-fit parameters were obtained by fixing
alternatively the $r_c$ and $S_0$ parameters to their 
best-fit values\footnote{Best-fit values and error bars are
computed using the python function {\tt curve\_fit}.}. 
This was repeated for different values of $n_{d}$.  }

\begin{figure*}
   \centering
    \includegraphics[scale=0.523]{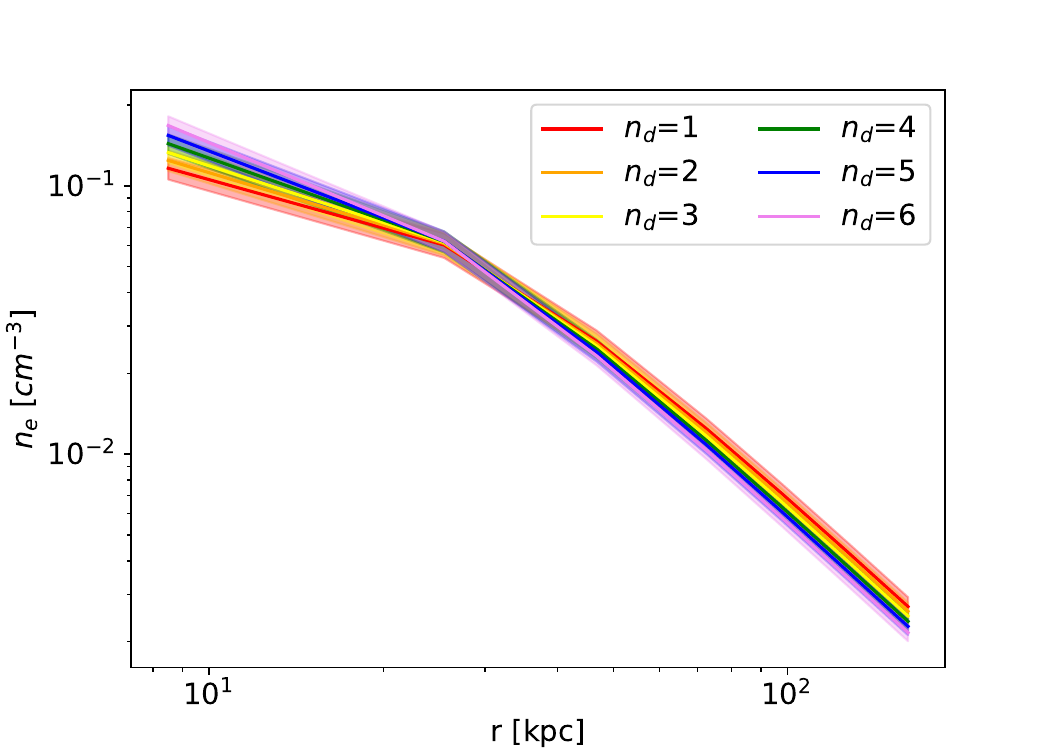}
   \includegraphics[scale=0.55]{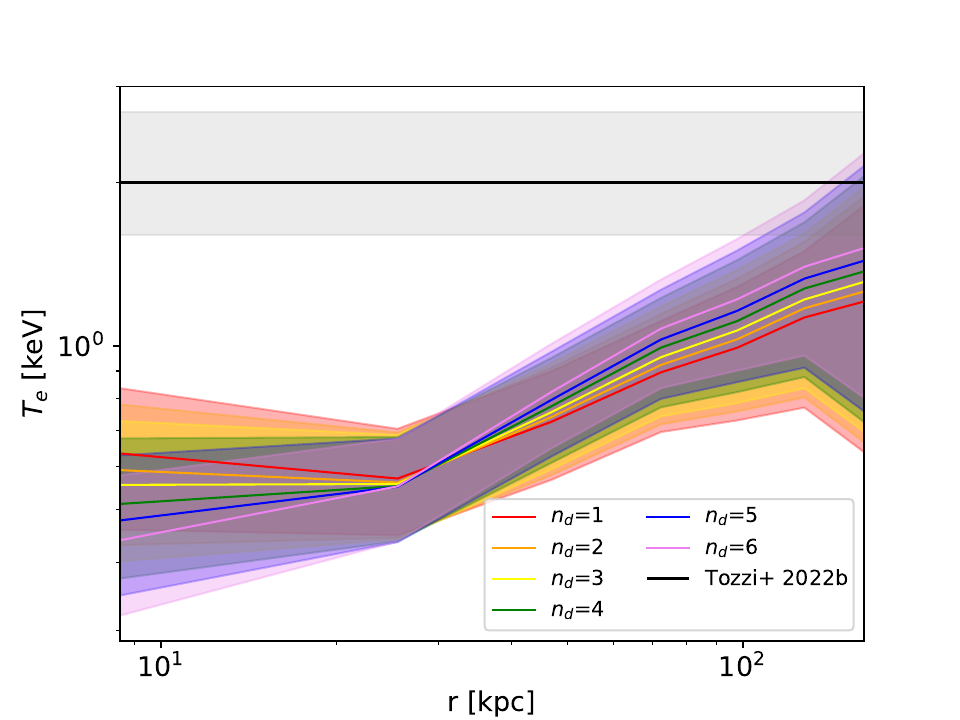}
   \caption{Electron density as a function of 
   the radius for different $n_{d}$ values (left). As we can see, 
   the profiles differ significantly only in the central $\sim 2$ arcsec 
   in which we apply the parametrization with $n_{d}$. Temperature profile as a function 
   of the distance from the central radio source for different $n_{d}$ values (right). 
   The solid black line represent the constant temperature value 
   of the ICM found by \cite{2022bTozzi}, assuming an isothermal 
   profile, while the shaded region mark the 1$\sigma$ errors 
   associated with the temperature.}
   \label{fig:density_temperature_profile}
\end{figure*}

Then, in order to obtain the 3D electron density profile, 
we calculated the values of $n_{e0}$ deprojecting the $\beta$-model,
{ using Equations 3 and 2 
and using the relation between the normalization of the 
global X-ray spectrum measured in 
\citet{2022bTozzi} and the electron density profile 
$n_e(r)$\footnote{See the XSPEC manual {\tt https://heasarc.gsfc.nasa.gov/xanadu/xspec/manual/node193.html}}}:

\begin{equation}
    norm=\frac{10^{-14}}{4\pi [D_{A}(1+z)]^{2}}\int n_{e}(r)n_{H}(r)dV ,
    \label{eq: normalization}
\end{equation}
where $D_{A}$ is the angular diameter distance, while $n_{e}$ and $n_{H}$ 
are the 3D density profiles of electrons and hydrogen atoms, respectively.{ For simplicity, the volume $dV$ is simply the 
spherical volume included in the spectral extraction region of 12 arcsec. This implies that we ignore the contribution beyond this radius, which is, in fact, negligible (as shown in Figure \ref{Spiderweb Circular Regions}) and
has a minor impact on the normalization.}
The values of $n_{e0}$ are reported in Table \ref{table:beta model parameters}.
The deprojected electron density profile is shown in the left panel of Figure \ref{fig:density_temperature_profile}. 
As expected, the 
profile does not show features associated with spatial scales 
and it is similar to a power law. The central value is parametrized
by the quantity $n_d$, while the best-fit profiles  are affected
overall by $n_d$ at a level lower than $5$\%. Despite the fact that the central value ($r<2$ arcsec) 
is not constrained, the density profile does not show the inner flattening 
of a regular core; instead, it is growing roughly as a power law with slope $\sim 1.7$ 
towards low radii, as expected when a cool core is present.

\subsection{Pressure and temperature profiles}
\label{subsec:temperature}

The inherent sensitivity limitation of the available X-ray data makes 
the measurement of a spatially resolved temperature profile almost impossible. 
An attempt to derive global temperature values from an X-ray spectral 
analysis in the annuli with radii of $2<r<7$ 
arcsec and $7<r<17$ arcsec has provided no indication of a difference in temperature 
in the two regions, with very loose contraints on both values.  Therefore, the 
only reliable indication of the global (emission-weighted) temperature of the 
proto-ICM is that obtained in \citet{2022bTozzi} from the X-ray spectrum, 
where we found $kT=2.0^{+0.7}_{-0.4}$ keV.

However, we can also constrain the temperature distribution in 
the Spiderweb proto-ICM by combining
the SZ pressure profile with the X-ray surface brightness distribution.
We remark that these two observables are completely independent and we do not make use
of the information provided by the X-ray spectral analysis. In addition, 
despite the fact that the thermal SZ spectral distortion is potentially affected by the temperature 
through relativistic corrections, the impact with respect to ALMA Bands~3 and 4 is expected to 
result in a systematic shift of the order of $\lesssim1.5\%$ of the 
total SZ signal \citep{2019Mroczkowski}, making this negligible compared to the 
intrinsic modeling and observational uncertainties (see \citealp{2023DiMascolo}). 
Similarly, the deprojection of the surface brightness from soft-band X-ray 
data is only mildly affected by the assumed value of the 
proto-ICM temperature.{ On the other hand, the normalization of the SZ
signal may be still affected by some systematics associated to the complex 
subtraction of the radio galaxy and the normalization of the surface brightness has some
dependence on the unknown metallicity of the ICM. Both of these aspects should be taken into account
when discussing our results.}

The radial profile of the temperature 
is directly obtained as the ratio of the pressure and electron density. 
In detail, the analytic estimate of the pressure distribution is obtained from the 
forward modeling of the ALMA data, while the binned density profile $n_e(r)$ 
is reconstructed from the soft-band X-ray surface brightness analysis 
(Section \ref{subsec:density_profile}). Since the radial annuli are determined
by the X-ray data, for consistency we post-processed 
the pressure profiles to estimate the $P_e(r)$ by 
marginalizing the values of the analytic realizations within each 
annulus (the resulting binned profiles are shown in Figure \ref{fig:pressure_profile}). 
Also in this case the profile can be
roughly approximated with a power law, but with an average slope of $\sim 1,  $ 
which is significantly flatter than the density profile. This 
directly implies a temperature gradient.

When simply assuming the ideal gas law, 
the average electron temperature within a given annulus is finally 
computed as the ratio $k_{B}T_{e}(r)=P_{e}(r)/n_e(r)$, as shown in the right panel of Figure \ref{fig:density_temperature_profile}. 
First, we notice a clear temperature 
gradient with value increasing from $\sim 0.6$ keV at $\sim 15$ kpc up to 
$\sim 1.8$ keV beyond 100 kpc. This result is interpreted as a very prominent cool core, 
showing the typical decrease by a factor of $\sim 3$ from the virial temperature 
reached at large radii and the emission-weighted temperature at the center. 
However, we do not see evidence for the expected flattening of the 
temperature profile with radius, probably because the X-ray signal 
disappear rapidly beyond 100 kpc (a distance still
below the $r_{500}$ value of $(220 \pm 30)$ kpc, as estimated in \citealt{2022bTozzi}). 

{ We note that the density-weighted temperature that we recovered from this profile, 
at radii larger than 2 arcsec, is $kT_{SZ,EW}= (0.7 \pm 0.3) $ keV; this value is 
lower by a factor of $\sim 3$ with respect to that obtained in 
\citet{2022bTozzi} with the X-ray spectral analysis ($kT=2.0^{+0.7}_{-0.4}$ keV, 
see the shaded area in Figure \ref{fig:density_temperature_profile}, right panel).  
A reason for this mismatch may be due to a residual contamination 
from non-thermal, diffuse emission
in the region used to extract the X-ray spectrum of the ICM.  
Indeed, a small non-thermal contamination may result
in a higher spectral temperature, while leaving the surface brightness 
practically unaffected. In addition, we also note that the
ICM metallicity is practically unconstrained.  The formal
best-fit value is $\sim 0.25 \, Z_\odot$ \citep[in term of][]{1989Anders}, 
but this may be much lower than the actual value, considering that the bulk of 
the emission is coming from the core that is expected to be highly enriched
by the ongoing SF. Assuming a value of $\sim 1\, Z_\odot$ ($\sim 2\, Z_\odot$)
would reduce the spectral temperature by 3\% (6\%) and the normalization of the 
density profile by 15\% (30\%), corresponding to an increase of the density-weighted temperature
by the same amount. This discussion suggests that the tension between the normalization of 
our temperature profile and the spectral value is significantly alleviated.}

{ On the other hand, the normalization of the SZ signal may be 
underestimated compared to the actual pressure distribution due to the impact of the large-scale interferometric filtering. The overall sensitivity from the available ACA+ALMA observations is, in fact, driven by the ALMA compact configuration measurements, with a maximum recoverable scale of $\sim 26~\mathrm{arcsec}$ (corresponding to $216~\mathrm{kpc}$ at the protocluster redshift). This implies that any diffuse SZ component (generally expected in the case of morphologically disturbed and forming structures) will be inherently suppressed, thereby resulting in a systematic reduction of the amplitude of the inferred pressure model. Similarly, non-thermal contributions to the overall pressure budget due to merger-induced turbulent motion of the forming ICM, for instant, would cause the SZ signal to deviate from the hydrostatic equilibrium expectation \citep{2022Bennett}. Finally, as already mentioned in \citet{2023DiMascolo}, the complex dynamical state of the Spiderweb protocluster might introduce a significant kinematic component to the total SZ effect observed in the direction of the system. The limited spectral coverage at millimetre bands, however, does not allow for  the different SZ contributions to be cleanly disentangled, resulting in a potential cross-contamination. In conclusion, the relatively low normalization of the temperature profile
inferred from the SZ signal may be associated to a systematic effect in the ACA+ALMA data analysis.}

To summarize, our results have revealed, for the first time,{ a clear temperature gradient associated with}
a cool core in a halo at $z>2$, showing that such structures form very 
rapidly as soon as the first cluster-sized halos start to collapse and virialize, 
around $z\sim 2.5$. The implication of these results will be discussed in the next sections.

\subsection{Entropy and cooling time profile}
\label{subsec:entropy}

{ In this section, we focus on the entropy and the cooling time profiles. 
If we assume the perfect gas law and consider a monatomic gas, 
the specific pseudo-entropy is given by:}

\begin{equation}
    K(r)\equiv kT(r)/n_{e}^{2/3}(r) = P_{e}(r)/n_{e}^{5/3}(r)\, 
    \label{entropy}
,\end{equation}

\noindent
{ where we combine the two independent measurements of $P_e(r)$ and $n_e(r)$.}

\begin{figure*}
   \centering
    \includegraphics[scale=0.51]{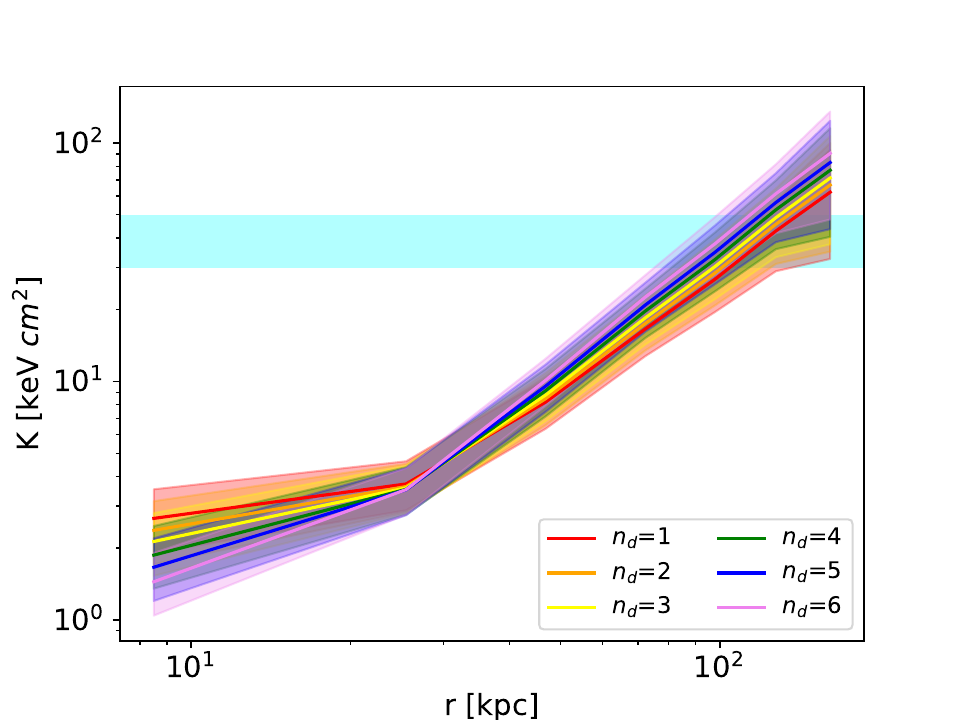}
   \includegraphics[scale=0.51]{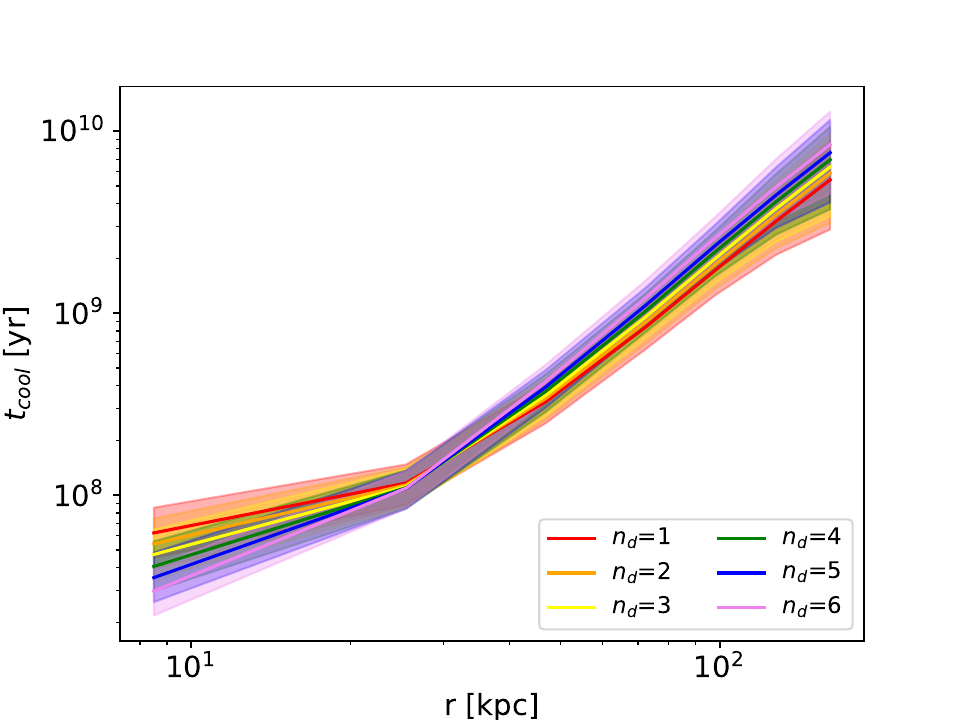}
   \caption{Entropy as a function of the distance 
   from the central radio source for different values of $n_{d}$ (left). {The cyan shaded area represents the threshold between cool core and non cool-core clusters found by \cite{2009Cavagnolo}.} According to this threshold, at radii $r<70$ kpc the Spiderweb can be classified as a strong cool-core cluster.
   Cooling time as a function of the distance 
   from the central radio source for different values of $n_{d}$, 
   assuming an ICM metallicity $Z=0.3 \rm Z_{\odot}$ (right).}
    \label{cooling time-entropy profile ERR}
\end{figure*}

In virialized clusters, the entropy of the ICM is observed to follow 
a self-similar radial distribution with some deviation at the center, 
where it flattens \citep{1999Ponman}. However, cooling may 
re-establish a power law behavior in this region \citep[see][]{2001Voit}. 
Outside the core, the entropy profile shows the typical power law 
generated by gravitational processes 
\citep[shocks and adiabatic compression,][]{2001Tozzi} as also found in 
hydrodynamical simulations.  For this reason, the entropy profile can be 
considered as the stratified thermal history of the ICM \citep{2005Voit}. 
The entropy profile obtained from the temperature and density profile 
according to Equation \ref{entropy} is shown in the left panel of 
Figure \ref{cooling time-entropy profile ERR}. Apart from the poorly constrained first bin, 
which nevertheless always shows a slight flattening for any reasonable 
value of $n_d$, the entropy profile follows a power law with a slope of 
$\sim 1.1$, as expected for cool core clusters.
If we consider the threshold between cool-core and non cool-core clusters, 
found to be between 30 and 50 
keV cm$^{2}$ by \cite{2009Cavagnolo}, the Spiderweb is classified as a strong cool-core cluster.

A very similar quantity is the cooling time, 
defined as the timescale in which the ICM entirely looses its internal 
energy through bremsstrahlung radiation. 
If we consider the ICM as a ionized gas in equilibrium, 
its internal energy is simply $(3/2)nkT$ and, therefore, the cooling time reads: 

\begin{equation}
    t_{\rm cool}\sim\frac{3}{2}\frac{nkT}{n_{e}n_{i}\Lambda(Z,T,n)}\, ,
    \label{tcool}
\end{equation}

\noindent
where $n_{e}$ and $n_{i}$ are the electron number density and 
the ion number density respectively, while $\Lambda(Z,T,n)$ is the 
cooling function which depends on the metallicity, $Z$, temperature, $T,$ and 
the number density \citep[see, e.g.,][]{2022DonahueVoit}. 
At high temperatures, the cooling function is 
dominated by bremsstrahlung emission, while at low temperatures ($kT\leq 2$ keV), 
the cooling function is significantly affected by line emission 
from heavy ions.{ Again, we used the relation $kT(r)=P_e(r)/n_e(r)$ in Equation \ref{tcool}.}
The cooling time profile is shown in the right panel of Figure 
\ref{cooling time-entropy profile ERR}.  
{ We assumed the cooling function of \citet{1993Sutherland}, 
with a metallicity of $Z=0.3 \, Z_{\odot}$ in units of \citet{1989Anders}}.
However,{ we note that (as expected) the uncertainty in the 
ICM metallicity is only mildly affecting the cooling time since it is degenerate with the 
emission measure $\propto n_e^2$, as discussed in Section 5.2.}
We find that the average cooling time is lower than 
one Gyr within 70 kpc. Usually, one Gyr is a reference value used to define a cool core
at $z\sim 0$ \citep[see][]{2010Hudson}, and it should be significantly shorter 
at $z\sim 2.2$ when the age of the universe is 3 Gyr. 
More physically motivated definitions of core radii have been
recently proposed \citep[][]{2023Wang}. According to our results, the cooling time
in the center can reach values below 0.1 Gyr, implying an unstable region
where cooling can rapidly occur via multiphase gas condensation and a CCA mechanism, 
possibly triggered by mechanical feedback. { In terms of CCA, we actually expect 
a higher frequency of triggering events at high-z, since diffuse baryons in the early universe 
are more often present in a multiphase stage and are more
disturbed, even without in situ condensation and SMBH feedback 
\citep[see][]{2012Gaspari,2013Gaspari}.}

\subsection{Total and ICM mass profile}
\label{subsec:massprofile}

\begin{figure*}
   \centering
    \includegraphics[scale=0.55]{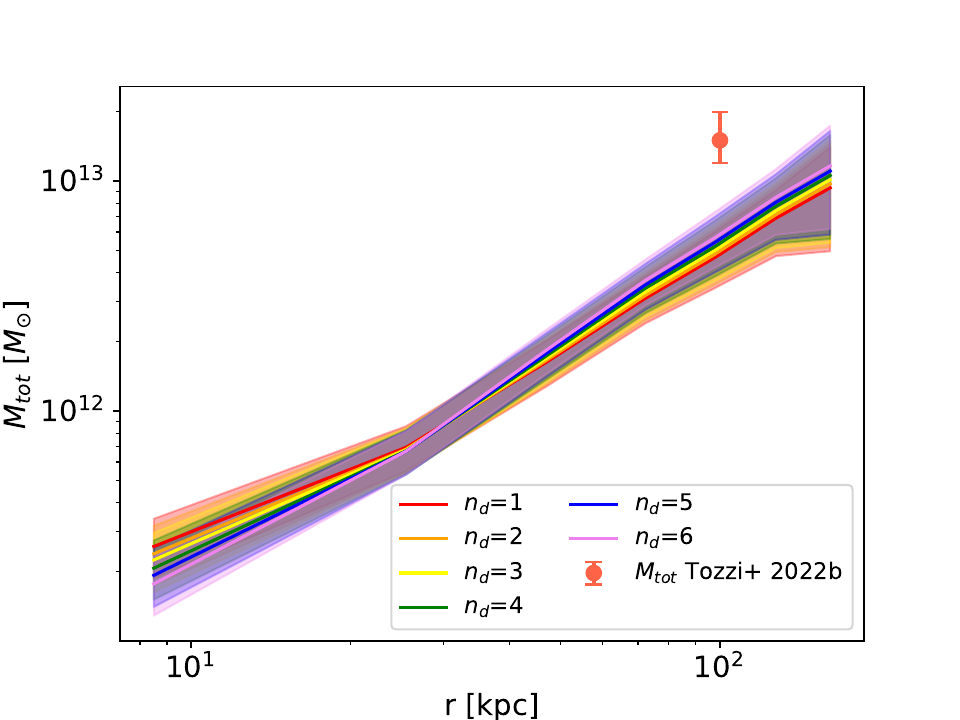}
    \includegraphics[scale=0.55]{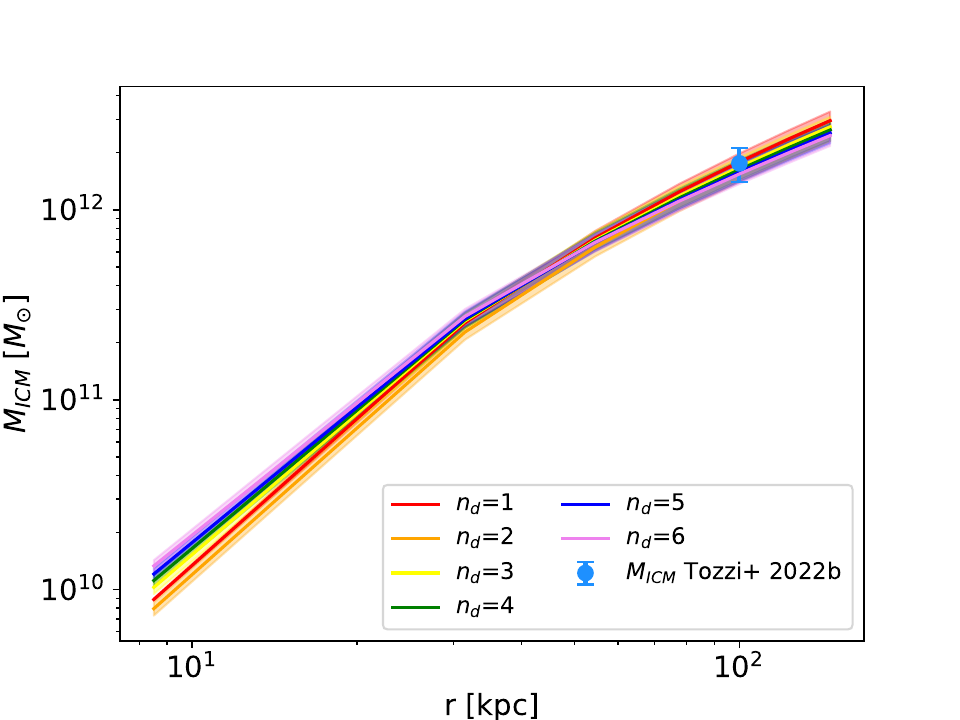}
   \caption{Total mass as a function of the radius 
   for different $n_{d}$ values (left). The profiles differ mostly in the 
   central $\sim 2$ arcsec. The red dot with errorbars represent the 
   value of total mass at $r=100$ kpc { based on the X-ray spectral 
   temperature} found by \cite{2022bTozzi}. 
   ICM total mass as a function of the radius 
   for different $n_{d}$ values (right). The blue dot with errorbars represents 
   the value of ICM total mass at $r=100$ kpc found by \cite{2022bTozzi}.}
    \label{fig:total_mass_profile}
\end{figure*}

Another quantity that we can immediately derive from the electron density 
and temperature profiles of the ICM is the total mass profile under the assumption 
of hydrostatic equilibrium, which is simply expressed as:

\begin{equation}
    M_{\rm tot}(<r)=-r\frac{kT(r)}{\mu m_{p}G}\biggr(\frac{dlog (\rho_{g})}{dlog(r)}+\frac{dlog(kT)}{dlog(r)}\biggr),
\end{equation}

\noindent
where $kT(r)$ is the temperature profile obtained in Section \ref{subsec:temperature}, 
$\rho_{g}$ is the gas density, $\mu$ is the mean molecular weight of the ICM, which 
is set to 0.6, and $m_{p}$ is the proton mass. 
The cumulative mass profile is shown in 
Figure \ref{fig:total_mass_profile} (left panel). {The total mass 
computed at $r\sim 100$ kpc ($M_{\rm tot}=(0.60\pm 0.15)\times 10^{13}\, 
\rm M_{\odot}$) is lower than
the value $M_{\rm tot}=(1.4\pm 0.3)\times 10^{13}\, \rm M_{\odot}$
obtained in \citet{2022bTozzi}, assuming a flat temperature profile. 
This discrepancy is entirely due to the difference between the X-ray spectral temperature
derived in \citet{2022bTozzi} and the temperature measured in this work. }

From the gas density $\rho_{g}(r)=n_{e}m_{p}A/Z$, where A and Z are the average nuclear charge and mass for the proto-ICM, we obtained the 
ICM cumulative mass profile, integrating over the volume:

\begin{equation}
    M_\textsc{icm}=\int_{0}^{r} \rho_{g}(r)dV \, .
\end{equation}

\noindent
The cumulative ICM mass profile is shown in 
Figure \ref{fig:total_mass_profile} (right panel).
We find that $M_\textsc{icm}=(1.64\pm0.19)\times10^{12}M_{\odot}$ within 100 kpc,  
in agreement with the average value of $(1.76\pm 0.30 \pm 0.17)\times10^{12}M_{\odot}$
(effectively an upper limit) found by \citet{2022bTozzi}, assuming a constant 
electron density distribution.

\section{Discussion}
\label{sec:discussion}

The analysis presented in this paper provides solid evidence 
for the presence of a cool core in the proto-ICM in the halo of the 
Spiderweb Galaxy and, at the same time, it offers some
evidence to support a depression in the ICM distribution in the quadrant
corresponding to the west radio jet. On the basis of the current data, 
we are not able to further constrain the properties of the proto-ICM, 
such as the actual MDR onto the BCG, or the energy 
budget of the feedback event -- or even, alternatively, the dynamical state
that may also be responsible for the asymmetry in the surface brightness distribution. 
 We explore the possible implications of our findings below.

\subsection{Estimate of the feedback energy budget}
\label{sec:energy_budget}

One of the results of this work is the presence of a possible depression 
in the X-ray surface brightness in the direction of the western jet, 
suggesting that the mechanical power of the central AGN may inflate 
a bubble of relativistic plasma in the ICM. For Fanaroff-Riley (FR) I radio sources 
in the Local Universe, the buoyancy arguments are often used to 
estimate the power that drives the expansion of the bubble 
(\citealt{1973Gull}, \citealt{2000Churazov}; see also 
\citealt{2012Fabian} and \citealt{2022HlavacekLarrondo} for reviews). 
Such estimates assume that during the subsonic expansion of the 
bubbles, buoyancy will eventually take over and drive 
the bubble up faster. For the subsonic expansion, 
most of the mechanical power of the outflow, $L_m$, goes into the 
enthalpy of the expanding bubble, according to the relation
\begin{eqnarray}
L_m\times t \approx \frac{\gamma}{\gamma-1}PV=\frac{\gamma}{\gamma-1}P\frac{4}{3}\pi r_b^3,
\label{eq:l_m}
\end{eqnarray}
where $t$ is the age of the bubble, $P$ is the ICM pressure, 
$V$ is the volume of a spherical bubble with radius, $r_b$, 
and $\gamma$ is the adiabatic index of the gas inside the bubble 
($\gamma=4/3$ for relativistic plasma). The corresponding  expansion velocity 
of the bubble can be approximated as:
\begin{eqnarray}
v_{\rm exp}=\frac{d r_b}{d t}\sim \frac{1}{3}\frac{r_b}{t}.
\label{eq:v_exp}
\end{eqnarray}

\noindent
On the other hand, the terminal velocity of a steadily rising 
low-density bubble is set by the balance of the buoyancy 
force, $F_b=\frac{4}{3}\pi r_b^3 \rho g$, and the drag force 
from the ICM  
acting on the bubble, $F_d=C_d \frac{1}{2}\pi \rho r_b^2 v_{\rm rise}^2$. Here, $\rho$ is the mass density of the ICM, 
$g$ is the gravitational acceleration in the cluster potential well, 
and $C_d$ is the drag coefficient. For a subsonic spherical bubble 
in an incompressible fluid with a large Reynolds number, $C_d\sim 0.5$;
while for flattened structures, it can be even greater 
\citep[e.g.,][]{2018Zhang}. Thus, the balance $F_b=F_d$ implies
\begin{eqnarray}
v_{\rm rise}=\left ( \frac{1}{C_d}\frac{8}{3} r_b g \right )^{1/2}=\left ( \frac{1}{C_d}\frac{8}{3} \frac{r_b}{R}\right )^{1/2}v_c ,
\label{eq:v_rise}
\end{eqnarray}
where we express the gravitational acceleration at distance, $R,$ 
from the center of the cluster in terms of the circular velocity as
$g=v_c^2/R$. Formally, this expression is valid as long as $r_b$ is much lower 
than the pressure scale height of the atmosphere. Once $v_{\rm rise}$ 
is larger than $v_{\rm exp}$, the buoyancy force deforms the bubble 
and moves it up in the cluster atmosphere. Requiring 
$v_{\rm rise} \approx v_{\rm exp}$ and solving for $t$ yields a 
lower limit on  the mechanical power
\begin{eqnarray}
L_m\gtrsim 3 \frac{\gamma}{\gamma-1}P\frac{4}{3}\pi r_b^2 \left ( \frac{1}{C_d}\frac{8}{3} \frac{r_b}{R}\right )^{1/2}v_c.
\end{eqnarray}
This corresponds to the approach used in \cite{2000Churazov} for the Perseus cluster. 
There, an additional explicit assumption was made that for Perseus 
(and many other FR~I sources) $r_b\sim R$ and, therefore, 
the term $\left ( r_b/R \right )^{1/2}$ can be omitted. Here, we relax this assumption  
and use the values for R and $r_b$ that we derived from our data. 
Adopting $R=60\,{\rm kpc}$, $r_b=11\,{\rm kpc}$, 
$P(R)=5\times 10^{-11}\,{\rm erg\,cm^{-3}}$, $\gamma=4/3$, $C_d=0.75$, 
and $v_c=700\,{\rm km\,s^{-1}}$ we obtained the estimate:
\begin{eqnarray}
L_m\gtrsim 10^{44}\,{\rm erg\, s^{-1}}.
\label{eq:lmec}
\end{eqnarray}

Various other variants of the mechanical power estimate used 
in the literature are compiled in \cite{2004Birzan}. They are also 
based on Eq.~(\ref{eq:l_m}), namely, 
$L_m=\frac{\gamma}{\gamma-1}PV/t$, where 
for the value of $t$, the following alternatives are considered: 
\\\\ (i) the sound crossing time from the cluster core to the bubble
\begin{eqnarray}
t_{cs}=\frac{R}{c_s},
\label{eq:t_cs}
\end{eqnarray}
where $c_s=\sqrt{\gamma_g\frac{kT}{\mu m_p} }$ is the sound speed in the thermal gas (ICM), $\gamma_g=5/3$, $kT=2\,{\rm keV}$, and $\mu=0.6$.
\\\\ (ii) The buoyant rise time to the current position of the bubble with the terminal velocity
\begin{eqnarray}
t_{\rm buoy}=\frac{R}{v_{\rm rise}},
\label{eq:t_buoy}
\end{eqnarray}
which is similar to what we previously estimated (see Eqs. ~\ref{eq:v_exp} and 
\ref{eq:v_rise}), but with a different normalization, and with $R$ replaced with $r_b$.  
\\\\ (iii) The {\sl refill} time is:\ 
\begin{eqnarray}
t_{\rm refill}=2\sqrt{\frac{r}{g}}=2\frac{\sqrt{rR}}{v_c},
\label{eq:t_refill}
\end{eqnarray}
 corresponding to the time required by a fluid element to cross the 
bubble's diameter starting with zero velocity and constant acceleration, $g$. 

Using the parameters we measured for the halo of the Spiderweb Galaxy, 
we obtained timescales of 80, 100, and 70~Myr for Eqs.~\ref{eq:t_cs}, \ref{eq:t_buoy}, 
and~\ref{eq:t_refill}, respectively. 
This corresponds to a mechanical power of the order of 
$L_m \sim 10^{43}\,{\rm erg\,s^{-1}}$ or larger.
Combining this outcome with the result of Eq.~\ref{eq:lmec}, 
we conclude that the  
AGN mechanical power is in the range of $\sim 10^{43}-10^{44}\,{\rm erg\,s^{-1}}$. 
{ This range of values is at the bottom of the distribution found by \citet{2015Hlavacek} in a sample 
of massive clusters out to $z\sim 0.7$. More precise measurements of cavity power in a 
sizeable sample of high-z clusters and protoclusters are needed to constrain the 
evolution of the average feedback power from BCGs. This is a clear science goal for 
future, high-resolution X-ray missions.}

We note that this estimate relies on the 
assumption that pressure inside the bubbles is comparable to the ICM pressure, 
so that the expansion of the bubble is subsonic 
and the contribution of the kinetic energy of the expanding shells 
around the bubbles can be ignored. In fact, the estimated 
non-thermal pressure in some regions of the jets is 
$P_{nt}\sim 10^{-9}\,{\rm dyne\,cm^{-2}}$. This is an order 
of magnitude larger than the average ICM thermal pressure, estimated as 
$P_{ICM}\sim 5\times 10^{-11}\,{\rm dyne\,cm^{-2}}$ \citep{2022Carilli}. 
If $P_{nt}$ is the actual pressure inside the cavities, 
then the energy deposited in the ICM will be higher than the previous estimates, 
while the age estimates will be correspondingly lower 
(see the value $t\sim 3\times 10^7\,{\rm yr}$ obtained in \citealt[][]{2022Carilli}). 
A much higher angular resolution would be needed to better characterizes the energy injection rate, 
by resolving the fine structure 
of the jets. In particular, the heads of the jets can propagate 
supersonically \citep[e.g.,][]{1974Blandford,1984Begelman} 
due to the momentum of a collimated flow, differently from the expansion of a cocoon. 
As noted by \cite{2022Carilli}, a part of the problem is that the Spiderweb 
Galaxy is a hybrid morphology radio source featuring a combination of 
FR~I and FR~II properties.  

To summarize, despite the large uncertainties, 
if we interpret the surface brightness depression as a cavity in the ICM carved by the jets, 
we obtain a result that states that the power 
of the mechanical feedback from the radio galaxy is 
comparable or larger than the total rest-frame 0.5-10~keV luminosity of the 
ICM, observed to be $2\pm 0.5 \times 10^{44}\,{\rm erg\,s^{-1}}$ \citep{2022bTozzi}.
This implies that the ongoing feedback is sufficient to compensate for ICM cooling losses.
Clearly, since there is no information on the time span over which the
feedback is acting, it is not possible to infer whether the cooling is already 
halted or, instead, whether cooling is steadily occurring while the feedback just kicked in. 
We note that stopping  the cooling is completely extremely difficult, as during the self-regulated 
cycle, the feeding flow usually occurs perpendicular to the feedback jets 
\citep[see maps in][]{2012Gaspari}.

 \subsection{The Spiderweb:\ A possible cool-core cluster in the making}
 \label{coolcore}

There are no sources at low redshift that can be directly compared to the central halo 
of the Spiderweb protocluster. 
In this work, we find that the electron density shows a very steep profile 
 that may reach $n_{e}\sim 10^{-1}$ $cm^{-3}$ in the central 10 kpc. 
In addition, we measure a very significant temperature gradient ranging from 
$kT\sim 0.6$ keV in the central 10 kpc to $\sim 2$ keV, clearly showing the presence 
of a strong cool core for the first time at $z>2$. 
These high density and low temperature values combined contribute to an extremely short
cooling time in the central regions, possibly lower than $t_{\rm cool}\sim 0.1$ Gyr. 
The main question here is whether the presence of a strong cool core
is the rule for massive virialized halos at $z>2$ or whether it is just
a rare occurrence.  Another key question is whether such 
high-redshift cool cores do host cooling flows with
significant MDRs, which is at variance with cool cores
in clusters at $z<1.5,$ where the cooling flows are generally quenched. 

We can address this last point by deriving the MDR in the case of 
a steady-state, isobaric cooling flow 
\citep{1994Fabian}. We recall that this is the so-called "classical cooling rate," 
{ which is  known to overestimate  the actual MDR upper limits
inferred in massive clusters from X-ray spectral analysis by10 to 100 times.} The advantage of 
assuming the classic isobaric cooling flow model is that 
the luminosity emitted in any temperature range $dT$ can be immediately written as:
\begin{equation}
    dL_{x}=\frac{5}{2}\frac{\dot{M}}{\mu m_{p}}k_{B} dT,
\end{equation}
where $\dot{M}$ is the MDR, $\mu$ is the mean atomic mass, 
$m_{p}$ is the proton mass, and $k_{B}$ is the Boltzmann constant 
\citep{1988Sarazin,2006Peterson}. To obtain the MDR, we calculated 
the X-ray luminosity in the soft band for the central $\sim 3$ arcsec 
(in which the cool core resides) as a function of the parameter, $n_{d}$. 
Accordingly, we get MDR values ranging between $\sim$ 250 $M_{\odot}/yr$ 
and $\sim$ 1000 $M_{\odot}/yr$, depending on the value of $n_{d}$. 
For $n_{d}=4$, which is the value that parameterizes the presence of 
a cool core, we get MDR$\sim 760\, \rm M_{\odot}$/yr. This value is close to 
the SFR values of $1390\pm 150\, \, \rm M_\odot$/yr 
\citep{2012Seymour} or $1150^{+1150}_{-580} \, \rm M_\odot$/yr
\citep{2013Rawlings} found using IR observations, implying that
a substantial fraction of the obscured SF may be sustained by
gas cooling directly out of the hot ICM halo.
Despite the fact that this is just a consistency argument and there are no additional proofs
for the presence of a cooling flow in the ICM, it is worth considering
this scenario in view of the potentially very young age of the 
cool-core/radio galaxy system in the Spiderweb protocluster. 
According to this scenario, the situation would be drastically different from that
in clusters at $z<1.5$, where the secular evolution of the BCG
regulates the ICM cooling and the residual SFR values observed in the 
BCG are typically more than one order of magnitude lower than the 
upper limits to the MDR measured in cool cores \citep[see][]{2016Molendi}
\footnote{However, it has been recently shown that
a MDR close to the isobaric values can still be accommodated in low 
redshift clusters when using a hidden cooling flow model \citep{2022Fabian}.}.

{ \subsection{Considering a possible high baryon fraction}
\label{fb}
Clusters and protoclusters are the most massive objects for which we can measure the baryonic mass (dominated by the ICM) and the total gravitating mass. Therefore, we can estimate the baryon fraction as the ratio of the ICM mass to the total cluster or protocluster mass.  
The baryon fraction we measure in this work at $r=100$ kpc is $f_B=0.33\pm 0.08$, 
about twice the value $f_B=0.16\pm 0.04$ found in \cite{2022bTozzi}. 
Therefore, we find a discrepancy at a level that is lower than 2$\sigma$.  
This is not due to the  measured ICM mass, whose values are
consistent in \citet{2022bTozzi} ($M_{ICM}= (1.76\pm 0.3\pm 0.17)\times 10^{12}\, M_\odot$)
and in this work ($M_{ICM}= (1.64\pm 0.19)\times 10^{12}\, M_\odot$).
In fact, the largest discrepancy with respect to our previous work is 
in the temperature and, hence, on the total mass measured at $100$ kpc. 
The total mass measured in \citet{2022bTozzi} from X-ray spectroscopy is
$M_{tot}=(1.4\pm 0.3)\times 10^{13} M_\odot$, while from the ratio of
X-ray emission and SZ signal we obtain $M_{tot}=(0.60\pm 0.15)\times 10^{13} M_\odot$. 

As we discuss in Sec. \ref{subsec:temperature}, an overestimation of the X-ray spectroscopic temperature
may be ascribed to a residual contribution of non-thermal
emission overlapped with what we consider thermal ICM emission. This unnoticed
contribution may have a significant impact in the spectrum, while having
just a minor effect on the surface brightness distribution. We
also mentioned in Section \ref{sec:morphology}, the possibility of an isotropic IC
emission due to a (unlikely) precession of the radio jets. 
At the same time, systematics in the derivation of the SZ signal, may be the cause
of the low temperature inferred from SZ, providing additional reasons to 
favor a total mass closer to the value of $M_{tot}=(1.4\pm 0.3)\times 10^{13} M_\odot$ 
measured in \citet{2022bTozzi}.

Nevertheless, we notice that both estimates of the baryon fraction are 
higher by a factor ranging from 2 to 5, 
than those measured in local clusters of comparable mass scale
\citep[see][]{2013Gonzalez}. Therefore, we argue that the
observed proto-ICM may be in an overdense phase 
and it has not reached virial equilibrium, yet with the underlying dark matter halo.
This may be due to an early phase of the gravitational collapse, 
or an inside-out shock heating of the diffuse
baryons.  As we know, in the same regions a significant amount of warm and cold
baryons have been observed.  Clearly, this is a highly
speculative scenario, that should be tested by observing
other proto-ICM cores and with deeper X-ray data, a task that became almost
impossible due to the dramatic loss of sensitivity of \emph{Chandra}. 
Given the uncertain significance of this tension, we did not further
explore this aspect within our data set.}

\subsection{  Weighing the potential commonality of  the Spiderweb protocluster}
\label{common}

If we focus on the association of an extreme starburst BCG with a cool core, 
the Spiderweb Galaxy has very few comparable sources in the $0<z<1.5$
range.  So far, we know that in this redshift range, 
almost all the relaxed and virialized clusters have a cool core
but no (or residual) cooling flows, as a result of the secular evolution 
of the nuclear activity in BCGs, and of the high duty cycle of radio AGN
in cluster cores.  Nevertheless, we know a few exceptions to this scenario, 
the most relevant case being the Phoenix cluster at z$\sim$0.6, that is classified as a 
rapidly evolving cool core cluster \citep{2013McDonald, 2015Tozzi, 2019McDonald, 2020Kitayama}. 
The Phoenix cluster has a much higher mass scale with respect to the Spiderweb protocluster, 
and, most importantly, it is observed at a cosmic epoch 5 Gyr after the Spiderweb. 
As such, the Phoenix must have experienced several feedback events that should have
quenched the potentially massive cooling flows. However, current estimates of the
MDR in the Phoenix amount to $350^{+250}_{-200} \, \rm M_\odot$/yr below 2 keV
\citep{2018Pinto}, comparable to the SFR in the BCG.  
In the case of the Phoenix, the AGN activity may be overcome by a high accretion rate
due to the large halo mass and an undisturbed dynamical evolution, leading to a 
cool core stronger than usual.  An alternative explanation is provided by a
young stage of the feedback process, a situation that is clearly preferred in the Spiderweb, 
due to the much younger cosmic epoch. A possible way out is to reconstruct the feedback 
history as recorded in the ICM, by identifying old cavities still buoyantly raising in the
outskirts, where the surface brightness of the ICM is much lower and the surface brightness
contrast of a cavity well below the sensitivity of current facilities. Another 
strategy to approach this issue is the statistical study of the halo population, searching 
for unbalanced cool cores at different cosmic epochs. 

While the occurrence of strong cool cores in halos at $z>2$ is unknown and it will be 
explored only with future X-ray and optical/IR/radio telescopes, we may argue that any massive cluster
at $z<1$ comparable to the Phoenix should be easily identified and characterized by
current X-ray facilities. While awaiting the results from the eROSITA survey, 
we may mention only three other cool core clusters at $z<1$ that may be 
compared to the Phoenix: ZwCl235 (z$\sim$0.08, \citealp{2023Ubertosi}, 
which also has a pair of cavities excavated by the central galaxy), 
RXC J2014.8-2430 (z=0.15, \citealp{2022Mroczkowski}), and 
RX J1720.1+2638 (z=0.16, \citealp{2023Perrott}). 
At $z>1$, we may include two dynamically relaxed cool core clusters:
SPT-CL J0607-4448 ($z=1.4$, \citealp{2023Masterson}) and SPT-CL J2215-3537 
($z=1.16$, \citealp{2023Calzadilla}). We argue that pushing the X-ray 
spectral analysis to the limits of current facilities in the Phoenix analogs 
at $z<1.5$ may shed light on the mechanism regulating the quenching and onset of 
cooling flows in cool cores, to be compared to what happens in younger halos at $z>2$.

Very little is known of the behavior of diffuse baryons in 
protoclusters. Both the presence of proto-ICM and the occurrence
of significant cooling events from the hot component is clearly expected to be 
limited to the most massive halos within the protoclusters themselves that, 
by definition, have an average overdensity about one order of magnitude lower
than in virialized halos. Still, cooling flows may form at any time, due to the short time scales that may prevent  feedback 
and the larger disturbances, as compared with the low-z counterparts.
We suggest, therefore, that the 
phase observed in the Spiderweb protocluster may not be an evolved cool core 
like the one observed in virialized clusters but, rather, a precursor phase
associated with the presence of the radio galaxy.

So far, the characterization of proto-ICM at $z>2$ has been extremely challenging, 
despite the fact that most of the protocluster have been identified thanks 
to the presence of a high redshift radio galaxy placed within an 
overdensity of galaxies. \cite{2007Venemans} found that $\sim$ 75$\%$ of radio 
galaxies are located within protoclusters traced by Ly$\alpha$ emitters (LAEs), 
considering a sample of nine radio galaxies 
\citep[see also][]{2009Matsuda,2012Galametz,2013Wylezalek}. 
Nevertheless, due to the extreme faintness of the expected diffuse X-ray emission, 
a deep and systematic X-ray surveys of such galaxies has never been attempted. 
A first step may be obtained by the X-ray follow-up of high-z radio 
galaxies with the most peculiar radio morphologies, such as 
the presence of a simultaneous FRI and FRII jets type \citep{2022Carilli}.
The first discoveries of these hybrid morphology radio sources 
date back to the work of \cite{2000GopalKrishna}, up to the most recent 
work of \cite{2023GopalKrishna}, with the discovery of radio galaxies 
4C+63.07 at z$\sim$ 0.3 and J1136-0328 at z$\sim$0.8. 
At higher redshift, instead, we mention the radio source 4C65.15 at $z=1.63$
\citep{2009MillerBrandt}, J1154+513 at $z=1.34$ and J1206+503 at $z=1.45$ \citep{2022Stroe}.

Alternatively, the presence of a protocluster may be obtained by the detection 
of LAE overdensity. 
Among the most promising sources, we take note of 
MRC 2104-242, which is a HzRG (z=2.49, \citealp{2001Overzier}) located within an overdensity of galaxies 
(about eight times the average field, \citealp{2014Cooke}) 
with typical SFR of $ \sim 10-100\, \rm M_{\odot}/yr$ and masses in the range of 
$\sim 10^{10}-10^{11}\, \rm M_{\odot}$. In addition, it is embedded 
within a Ly$\alpha$ emitting halo that extends for $\sim$ 136 kpc, 
showing two large clumps and a very large velocity distribution 
($\sim 1000-5000$ km/s, \citealp{2001Overzier}).
We also mention the radio galaxy 4C-00.62 \citep{2006Kajisawa} which is located 
at $z=2.53$ within the USS 1558-003 protocluster, with a LAE overdensity 
showing three prominent structures \citep{2018Shimakawa}.

At lower redshift, 7C 1756+6520 ($z=1.42$, \citealp{2009Galametz, 2010Galametz}) 
inhabits an environment very similar to the Spiderweb, 
with a prominent galaxy overdensity (among which 21 spectroscopically confirmed)
with several AGN candidates. The protocluster found in the field of a quasar 
at z$\sim$6, SDSS J1030+0524 \cite{2019Gilli} hosts a radio galaxy 
surrounded by an overdensity of galaxies 
(11 spectroscopic members in a distance of $\sim$ 1.15 Mpc; \citealp{2020Damato}) 
with $SFR \sim 8-600\, \rm M_{\odot}/yr$. This structure seems far from being 
virialized but shows diffuse X-ray emission that
can be at least partially associated with an expanding bubble of gas at temperature 
T$\sim$ 5 keV \citep{2019Gilli}. In this framework, 
the presence of promising targets coupled to the lack
of a systematic approach and of a high-resolution, wide-angle X-ray survey, force us
to conclude that the best effort at the moment is to develop a systematic approach with SZ observations and to exploit, at the same time, current X-ray facilities 
to push as deeply as possible on the best-candidate  protoclusters identified so far.

\section{Future perspectives}
\label{sec:Future}

Our results demonstrate the central role of a sharp \emph{Chandra}-like PSF 
for detailed analyses of the ICM properties of high-z protoclusters. 
The role of X-ray data with high-angular resolution will be key for the extension 
of this approach to a large sample of protoclusters. 
Currently, \emph{Chandra} is the only facility that can provide data of 
the required quality.  As a complementary tool, XMM-Newton can contribute with the
characterization of the X-ray properties of high-z radio galaxies, but with 
a loose grasp on possible extended thermal emission on scales of $\sim 10$ arcsec. 
Therefore, the future of this science case heavily depends on the 
future of high-resolution X-ray astronomy, carried out with 
the advent of the next generation of X-ray satellites, such as 
AXIS \citep{2019Mushotzky, 2020Marchesi}, the survey and time domain mission 
STAR-X \citep{2022Zhang}, Lynx \citep{2018LynxTeam}, and Line Emission
Mapper  (LEM; \citealp{2023Kraft}). Thanks 
to a large field of view to increase the discovery space, 
a high resolution (in the range 0.5-3 arcsec) and a high effective area, particularly
in the soft band,
it will be possible to characterize the emission of the ICM at $z\sim 3$ 
and untangle this emission from the AGNs emission, to map density, temperature, 
and also non-thermal emission at physical scales below 5 kpc.

As already discussed in Section \ref{sec:Introduction}, the complementary and 
alternative way to study protoclusters and the formation, evolution and 
structure of the ICM, is provided by interferometers and single dish 
instruments from the ground.  In the last decade, interferometers like ALMA and 
ACA have been able to observe the diffuse gas and confirm the membership in 
protoclusters at $z>2$ using molecular gas emission line such as CO lines.
Unfortunately, current (sub-) millimeters interferometers do not allow us to explore 
significantly larger areas without being exceedingly time-consuming due to a small 
field of view. Also, they are only sensitive to emissions on scales 
inversely proportional to their baseline separations, making them insensitive 
to large-scale emission (the so-called ``missing flux" problem). 
For these reasons they have extreme limitations for mapping the tenuous gas 
in protoclusters and the SZ effect on larger scales with a 
high angular resolution. 

Recently, a promising improvement that has been made on ALMA 
is the introduction of Band 1 (35-50 GHz), which allows us to increase 
the maximum recoverable scale, making it easier to study 
the SZ signal from diffuse gas on larger scales. This will also allow us 
to provide unprecedented detailed analysis of the ICM and its thermodynamics 
and to detect substructures in protoclusters. Unfortunately, 
at this point interferometers are and will be mainly useful to focus on 
known targets for detailed follow-up studies, with very little room for
the discovery of new targets. This situation leads to the problem of 
"source starvation."

Single-dish facilities, on the other hand, should allow us to map 
this emission with greater resolution and at higher sensitivities. 
One possible future single dish that would combine dish diameters and with a 
large field of view is AtLAST \citep{2020Klaassen,2022Ramasawmy}. Indeed, 
AtLAST's key parameters include a 50 m aperture and a 2 deg field of view.  It will be built in a location with excellent atmospheric transmission 
at millimeter/submillimeter wavelengths. Thus, the combination of arcsec-scale 
angular resolution with broad frequency coverage ($\approx$30-950~GHz) and 
multiple bands sampling the SZ effect spectrum, will allow us 
to study the hot ICM on larger scales, sample the resolved SZ effect 
across a wide range of redshifts and in different environments 
(from groups to clusters and protoclusters), detect dynamical effects 
on diffuse gas and explore variations in pressure and temperature profiles. 
Considering that both AtLAST and STAR-X may be operating in 2030, the perspective
of having a systematic and statistically significant investigation 
of the proto-ICM in the range $2<z<3$ may become reality in less than ten years.

\section{Conclusions}
\label{sec:Conclusions}

In this paper, we present a detailed analysis of the thermal diffuse emission 
of the proto-ICM detected in the halo of the Spiderweb Galaxy at 
$z=2.16$ within a radius of $\sim 150$ kpc.
We  simultaneously analyzed the deep X-ray data from {\sl Chandra} and 
the SZ signal obtained with ALMA sub-millimeter data. 
Our results can be summarized as follows:

\begin{itemize}

\item  Thanks to the \emph{Chandra} X-ray data we find that the 
azimuthally averaged surface brightness is well characterized in the
soft band, while the photometry in the hard band is 
consistent with no detectable signal. This confirms that the diffuse emission 
is consistent with being bremsstrahlung emission at $\sim 2$ keV or lower. 

\item { We also find a mild evidence of asymmetry in the X-ray surface brightness, 
in the form of a depression in the sector corresponding to the 
direction of the western jet.  This feature may be tentatively
associated with a X-ray  cavity carved into the hot baryons and
filled with low-density relativistic plasma, which originated as a 
displacement of gas due to the radio jets. In this case, 
the estimated energy power 
of the associated feedback events is in the range $10^{43}-10^{44}$ erg/s, depending on the method used
to estimate the mechanical power. Alternatively, the asymmetry 
may be due to the young dynamical status of the halo. } 

\item From the 2D analysis of the X-ray data, we 
find that the bulk of the X-ray emission (excluding the central core within a radius of
3 arcsec) has the centroid located at a distance 
of $\sim 2.3$ arcsec from the AGN. This shift is { along the same
direction of} the SZ effect offset (6.2 $\pm$ 1.3 arcsec) measured with ALMA. 

\item We combine the electron density profile from X-ray data and 
the pressure profile from the SZ effect to derive for the first time the 
temperature profile of the proto-ICM. We find a strong temperature gradient,
corresponding to the increasing electron density profile towards 
the central AGN, revealing the presence of a strong cool core. 
{ This result is robust despite the $2 \sigma$ disagreement at face value between 
the normalization of the temperature profile and the X-ray spectral, single temperature 
obtained in \cite{2022bTozzi}.}

\item {From the temperature and electron density profiles, 
we obtain the total mass profile assuming hydrostatic equilibrium, as well as 
the ICM mass profile. We find a total mass value of 
$M_{\rm tot}(100 kpc)=(0.60\pm 0.15)\times10^{13}$$M_{\odot}$, a factor of 2
lower than the value measured in \cite{2022bTozzi} assuming an isothermal ICM of $\sim 2 $ keV.
The ICM mass is measured to be $M_\textsc{icm}=(1.64\pm 0.19)\times 10^{12}\, \rm M_{\odot}$, 
in agreement with the upper limit of $(1.76\pm0.30\pm 0.17) \times 10^{12}\, \rm M_{\odot}$ 
found by \cite{2022bTozzi} using a constant density profile.}

\item Using the temperature and electron density profile, 
we obtain the entropy profile, that follows a power law as expected in cool cores. 
Correspondingly, the cooling time profile shows values lower than 
1 Gyr below 70 kpc from the central AGN, possibly reaching values as low as 0.1 Gyr in the 
innermost 20 kpc, suggesting  that the cooling may occur on a time scale shorter than
the AGN duty cycle.  

\item We argue that, if the feedback is not effective yet in reheating the cooling ICM, 
a cooling flow may develop, with a MDR {in the range 250-1000 $M_{\odot}/yr$, consistent with}
the SFR values found using IR observations, implying that
a substantial fraction of the obscured SF may be sustained by
gas cooling directly out of the hot ICM halo.

\end{itemize}

To summarize, for the first time we constrain the thermodynamic properties 
and the hot baryon circulation 
in the core of a protocluster at $z>2$, thanks to an unprecedented 
combination of deep, high-resolution {\sl Chandra} X-ray data and SZ ALMA data. 
As expected, the first stages of formation of protoclusters are 
characterized by an extreme environment with energetic feedback events 
that may coexist for a relatively short time with massive cooling flows episodes. 
The systematic studies of proto-ICM halos in protoclusters may reveal a new, short
and intense phase of heating and cooling, just before the BCG, its AGN activity and 
the ICM are driven into the secular evolution { achieved} when the feedback efficiently quench
cooling flows, as{ commonly} observed at $z<1.5$.

\begin{acknowledgements}
We thank the anonymous referee for detailed comments and positive criticism 
that helped us improving the quality of this manuscript. 
M.L.\ and P.T.\ acknowledge financial contribution from the PRIN MIUR 2017
"Zooming into Dark Matter and proto-Galaxies with Cosmic Telescopes". 
L.D.M.\ is supported by the ERC-StG ``ClustersXCosmo'' grant agreement 716762. 
L.D.M.\ further acknowledges financial contribution from the agreement ASI-INAF n.2017-14-H.0.
A.L.\ acknowledges financial support from the European Research Council (ERC) 
Consolidator Grant under the European Union's Horizon 2020 research and 
innovation programme (grant agreement CoG DarkQuest No 101002585). 
M.G.\ acknowledges partial support by NASA HST GO- 15890.020/023-A, and the \textit{BlackHoleWeather} program.
\end{acknowledgements}
      

\bibliographystyle{aa}
\bibliography{Spiderweb_analisi_proto_ICM.bib}

\appendix

\section{Pressure models}\label{app:sz}
To date, we continue to lack an exhaustive theoretical description of the expected distribution of the thermodynamic properties of the ICM forming within protocluster structures. To fully account for potential biases and variance introduced when forward modeling the proto-ICM thermodynamic profiles by means of models calibrated on low-redshift cluster samples, \citet{2023DiMascolo} performed a comparative analysis of the SZ signal in the direction of the Spiderweb protocluster by means of different gNFW models from the literature, namely: the pressure profile derived by \citet{2010Arnaud}, using both its universal formulation (A10 UP) and the one constrained on the subset of morphologically disturbed (A10 MD) clusters; the universal (M14 UP) and non-cool-core (M14 NCC) pressure profiles by \citet{2014McDonald}; the median profiles derived from the simulated cluster set in \citet{2015LeBrun} and under different assumptions for the AGN contributions (L15 REF, L15 8.0, L15 8.5; we refer to \citealt{2014LeBrun,2015LeBrun} for details on the physical prescriptions for the AGN-driven heating); the self-similar extended pressure model by \citet{2017Gupta}.

In this work, we repeat the same analysis but fixing the coordinates of the SZ centroid to the one inferred from the X-ray analysis in Section \ref{sec:combined}. In such a way, the parameter set for each parameter reduces to the mass term only. A summary of the best-fit values of the virial mass $M_{500}$ 
for each pressure model is reported in Table~\ref{app:tab:sz}.

\begin{table}[]
    \renewcommand{\arraystretch}{1.5}
    \centering
    \caption{Comparison of the best-fit parameters for the different SZ models employed in the analysis. }
    \begin{tabular}{lcc}
        \hline\hline
    &  $M_{500}$ & ref. \\
    & [$10^{13}~\mathrm{M_{\odot}}$] &  \\
        \hline
        A10 UP  &  $2.47\substack{+0.44\\-0.54}$ & \citealt{2010Arnaud}   \\
        A10 MD  &  $2.01\substack{+0.39\\-0.49}$ & \citealt{2010Arnaud}   \\
        M14 UP  &  $2.82\substack{+0.53\\-0.65}$ & \citealt{2014McDonald} \\
        M14 NCC &  $2.84\substack{+0.53\\-0.66}$ & \citealt{2014McDonald} \\
        L15 REF &  $2.59\substack{+0.42\\-0.48}$ & \citealt{2015LeBrun}   \\
        L15 8.0 &  $4.30\substack{+0.55\\-0.68}$ & \citealt{2015LeBrun}   \\
        L15 8.5 &  $6.54\substack{+0.78\\-0.95}$ & \citealt{2015LeBrun}   \\
        G17 EXT &  $2.20\substack{+0.39\\-0.55}$ & \citealt{2017Gupta}    \\
        \hline
    \end{tabular}
    \tablefoot{The parameter $M_{500}$ denotes the mass enclosed within the spherical volume of radius $r_{500}$ with an average density equal to $500\times$ the critical density of the Universe at the redshift $z$ of the system.}
    \label{app:tab:sz}
\end{table}

\end{document}